\documentclass{emulateapj}
\usepackage{graphicx}
\usepackage{bm}
\usepackage{amssymb}
\usepackage{amsmath}
\usepackage{cancel}
\usepackage{hyperref}

\usepackage{epstopdf}
\usepackage{natbib}
\usepackage{epsfig}
\usepackage{threeparttable}

\newcommand{\Ang}{\mbox{ \AA}}     

\newcommand{\avg}[1]{\ensuremath{\langle #1 \rangle}}
\newcommand{\bma}{\begin{math}}
\newcommand{\ema}{\end{math}}
\newcommand{\beq}{\begin{equation}}
\newcommand{\eeq}{\end{equation}}
\newcommand{\beqa}{\begin{eqnarray}}
\newcommand{\eeqa}{\end{eqnarray}}
\newcommand{\bc}{\begin{center}}
\newcommand{\ec}{\end{center}} 
\newcommand{\bit}{\begin{itemize}}
\newcommand{\eit}{\end{itemize}}

\usepackage{color}

\font\BFd=cmmib10
\font\BFt=cmmib10
\font\BFs=cmmib10 scaled 700
\font\BFss=cmmib10 scaled 500

\def\bbox#1{%
\relax\ifmmode
\mathchoice
{{\hbox{\BFd #1}}}
{{\hbox{\BFt #1}}}
{{\hbox{\BFs #1}}}
{{\hbox{\BFss #1}}}
\else \mbox{#1} \fi }

\begin{document}

\title{Future constraints on the reionization history and the ionizing sources from Gamma-ray Burst Afterglows}
\author{Adam Lidz$^{1}$, Tzu-Ching Chang$^{2,3}$, Llu\'is Mas-Ribas$^{2,3}$, Guochao Sun$^{3}$}
\altaffiltext{1} {Department of Physics \& Astronomy, University of Pennsylvania, 209 South 33rd Street, Philadelphia, PA 19104, USA}
\altaffiltext{2}{Jet Propulsion Laboratory, California Institute of Technology, 4800 Oak Grove Dr., Pasadena, CA 91109, USA}
\altaffiltext{3}{California Institute of Technology, 1200 E. California Blvd., Pasadena, CA 91125, USA}
\email{alidz@sas.upenn.edu}

\begin{abstract}
We forecast the reionization history constraints, inferred from Lyman-alpha damping wing absorption features, for a future sample of $\sim 20$ $z \geq 6$ gamma-ray burst (GRB) afterglows. We describe each afterglow spectrum by a three-parameter model. First, $L$ characterizes the size of the ionized region (the ``bubble size'') around a GRB host halo. 
Second, $\langle{x_{\rm HI}\rangle}$ is the volume-averaged neutral fraction outside of the ionized bubble around the GRB, which is approximated as spatially uniform. Finally, $N_{\mathrm{HI}}$ denotes the column-density of  a local damped Lyman-$\alpha$ absorber (DLA) associated with the GRB host galaxy.  The size distribution of ionized regions is extracted from a numerical simulation of reionization, and evolves strongly across the Epoch of Reionization (EoR). The model DLA column densities follow the empirical distribution determined from current GRB afterglow spectra. We use a Fisher matrix formalism to forecast 
the $\langle{x_{\rm HI}(z)\rangle}$ constraints that can be obtained from follow-up spectroscopy of afterglows with SNR = 20 per $R=3,000$ resolution element at the continuum. We find that the neutral fraction may be determined to better than 10-15\% (1-$\sigma$) accuracy from this data across multiple independent redshift bins at $z \sim 6-10$, spanning much of the EoR, although the precision degrades somewhat near the end of reionization. A more futuristic survey with $80$ GRB afterglows at $z \geq 6$ can improve the precision here by a factor of $2$ and extend measurements out to $z \sim 14$.
We further discuss how these constraints may be combined with estimates of the escape fraction of ionizing photons, derived from the DLA column density distribution towards GRBs extracted at slightly lower redshift, $z \sim 5$. This combination will help in testing whether we have an accurate census of the sources that reionized the universe.  
\end{abstract}


\section{Introduction} \label{sec:intro}

Gamma ray bursts (GRBs) are, electromagnetically, the brightest explosions in the universe. The long-duration class of GRBs are thought to be powered by the collapse of massive rotating stars \citep{Woosley93}: as nuclear fuel is exhausted at the end of the star's life, the core loses pressure support and implodes to form a black hole. In this process, jets of energetic particles are produced, which plow through the remaining stellar envelope. This relativistic outflow leads to highly-beamed gamma-ray emission, while subsequent afterglows occur at longer wavelengths as the ejecta runs into surrounding gas in the circumburst and interstellar media (ISM). The incredible luminosity of these events makes long-duration GRBs detectable from even some of the earliest phases of cosmic history. 
Indeed, long-duration GRBs have been observed back into the Epoch of Reionization (EoR) \citep{Tanvir_2018GRB7p8,Melandri_2015,Chornock:2013una,Cucchiara11,Salvaterra2009,Tanvir2009,Totani:2005ng}, when early generations of galaxies and/or accreting black holes emitted ionizing photons and gradually filled the universe with ionized gas (e.g., \citealt{Loeb2013}). Therefore, GRB afterglows can be exploited as bright backlights, with follow-up spectroscopy of the afterglows at high spectral resolution and signal-to-noise ratio providing valuable constraints on the neutral fraction of the intergalactic medium (IGM) (e.g., \citealt{Totani:2005ng,Hartoog:2014jda}) and the chemical enrichment history of the ISM in the GRB host galaxy (e.g., \citealt{Lamb:1999qv,Savaglio_2006NJPh}). 

A number of projects are being proposed to study reionization and metal enrichment using GRB afterglows. These include the Gamow Explorer which is expected to find roughly $\sim 20$ GRB afterglows in the EoR at $z \geq 6$ using an X-ray telescope with a wide field of view, exploiting that the GRB emission at high redshifts is largely shifted into the X-ray band  \citep{White2020,Ghirlanda2015}. The Gamow Explorer will then determine photometric redshifts for these sources, using an on-board infrared telescope, from the drop in flux blueward of Lyman-alpha (Ly-$\alpha$) at the source redshift. This will enable rapid targeted follow-up spectroscopic campaigns using the James Webb Space Telescope (JWST) in space or existing eight-meter and future thirty-meter class telescopes from the ground. The Gamow Explorer is proposed to launch in 2028, with a nominal three year mission lifetime. The detectable GRB rate during the EoR from the Gamow Explorer should improve on the Swift telescope's \citep{Gehrels2004} detection rate of such events by roughly an order of magnitude. In addition, HiZ-GUNDAM, which is similar in scope and timescale to the Gamow Explorer, is being planned in Japan \citep{Yonetoku:2014bga}.
In the slightly longer term, a larger European-led satellite, Theseus, is being proposed for launch in 2032, and should detect between 30-80 $z \geq 6$ GRBs \citep{Amati:2017npy,Theseus2018,Amati2021}. 

Motivated by these future prospects, the aim of the current paper is to forecast the reionization history constraints expected from these upcoming missions. The key spectral signature of remaining neutral gas in the IGM is the damping wing of the Ly-$\alpha$ line owing to its natural line width \citep{MiraldaEscude:1997qb}. This leads to spectrally extended absorption redward of Ly-$\alpha$ at the source redshift, provided significantly neutral gas -- with a neutral hydrogen fraction of order unity -- remains in the IGM at the GRB redshift. 

In this regard, GRB afterglows offer a few potential advantages over quasars \citep{MiraldaEscude:1997qb,Barkana:2003ja} which have nevertheless provided our best backlights for studying the IGM thus far. First, the intrinsic unabsorbed spectrum of a GRB afterglow has a simple power-law form around Ly-$\alpha$ at the source redshift. In contrast, in the case of quasars, the damping wing signature needs to be extracted in the presence of a Ly-$\alpha$ emission line (e.g., \citealt{Davies:2018yfc}). Second, the quasars themselves can ionize significant regions of neutral hydrogen around them. Furthermore, these sources likely reside in highly biased locations where surrounding galaxies form earlier than in typical regions of the universe. Reionization may hence occur earlier in quasar environments than in average parts of the universe \citep{Lidz:2007mz}. Together, the ionizing impact of quasars and surrounding galaxies may lead to weaker damping wing signatures around quasars than around less-biased sources such as long duration GRBs, which are expected to trace more typical regions of massive star formation. This may make the damping wing signature both more difficult to detect around quasar sources, and also complicate the interpretation of quasar-based measurements which require modeling the relationship between the quasar environments and typical parts of the IGM. Third, the quasar luminosity function falls very steeply with increasing redshift and current forecasts suggest that only a single quasar (with UV absolute magnitude $M_{\mathrm{AB}} \leq -26$) is detectable in the observable universe at $z \gtrsim 9$ \citep{Fan2019}. GRB afterglows provide the best candidates for observing damping wing signatures in the early phases of the EoR.

The GRB afterglow measurements are not without their challenges, however. First, one needs to detect the GRBs and their afterglows, rapidly determine photometric redshifts, and obtain high quality follow-up spectroscopic observations of the fading afterglow. Second, current samples of GRB afterglows frequently show strong damped Ly-$\alpha$ (DLA) absorption from neutral hydrogen in the ISM of the host galaxy (e.g., \citealt{Tanvir:2018pbq}). In some cases, the host DLA can swamp the signature of damping wing absorption from neutral gas in the IGM \citep{MiraldaEscude:1997qb}. Third, a challenge for interpreting damping wing signatures of the IGM towards both GRB afterglows and quasar backlights is the patchiness of the reionization process \citep{McQuinn:2007gm,Mesinger:2007kd}. This leads to a significant sightline-to-sightline scatter in the neutral fraction towards these background sources. More importantly, it implies that the damping wing profile is at least partly sensitive to the precise distribution of neutral gas along each line of sight and it is unclear how to best model the resulting spectra. 

Here we consider a simple parameterization (see also, e.g., \citealt{McQuinn:2007gm}) that partly accounts for the patchiness of the reionzation process, includes the plausible impact of host DLAs, and allows for rapid, fully analytic forecasts of the reionization history constraints obtainable from future GRB afterglow spectral samples. At each redshift, afterglow spectra are described by a bubble size parameter, $L$, a neutral fraction, $\avg{x_{\rm{HI}}}$  (approximated as uniform outside the host bubble), and a host DLA column density, $N_{\rm{HI}}$. The bubble size and column densities are drawn from probability distributions, inferred respectively from reionization simulations and observations, to allow for the likely broad sightline-to-sightline scatter in these quantities. In this framework, we use the Fisher matrix formalism to determine the neutral fraction constraints across several redshift bins, marginalized over the bubble size and column density parameters. 

In addition to providing backlights for studying the IGM damping wings, GRBs also provide valuable information about the escape fraction of ionizing photons \citep{Chen:2007wi,Tanvir:2018pbq} and the star-formation rate density (SFRD) at high redshifts. Specifically, a plausible working assumption is that GRB afterglows provide lines of sight towards massive stars which are representative of the pathways traversed by ionizing photons, as they propagate out into the IGM from the sources of reionization. The fraction of afterglows that have host columns less than roughly $\lesssim 10^{18}$ cm$^{-2}$, corresponding to Lyman continuum optical depths at the photoionization threshold of $\tau_{{\rm HI}} \lesssim 6.3$, then provides an estimate of the escape fraction.\footnote{This is just a rough estimate. A more detailed determination of the escape fraction computes $\avg{\rm{exp}(-\tau_{\rm HI})}$, averaged over the sightlines in the sample and over frequencies above the photoionization threshold, for some assumed ionizing spectral shape.}
Further, if the conversion between GRB rate and SFRD can be calibrated accurately at lower redshift, the redshift evolution of the GRB rate can be used to determine the SFRD during the EoR, including contributions from lower mass host galaxies that are challenging to observe directly with other star formation indicators (e.g., \citealt{Blain2000,Kistler:2007ud,Robertson2012,Vergani2015}). Measurements of the neutral fraction can be combined with estimates of the escape fraction and internal or external determinations of the SFRD to address whether the neutral fraction constraints are consistent with known source populations. We consider this in the context of our Fisher matrix analyses.

The outline of this paper is as follows. In \S \ref{sec:method} we review the IGM damping wing signature and the impact of DLA host absorbers. We then describe the key ingredients of our model: the redshift distribution of GRB afterglows (\S \ref{sec:GRB_zdist}), the reionization history model (\S \ref{sec:fiducial_history}), the bubble size distribution (\S \ref{sec:bubble_dist}), and the probability distribution of host column densities (\S \ref{sec:dlas}). \S \ref{sec:fisher} describes our Fisher matrix calculations and characterizes parameter degeneracies, while \S \ref{sec:forecasts} forecasts the resulting neutral fraction constraints. In this section we further discuss the dependence of these constraints on the quality of follow-up spectroscopic observations and the possibilities for obtaining the requisite spectra. \S \ref{sec:census} discusses the prospects for determining the escape fraction of ionizing photons and the resulting implications. We conclude in \S \ref{sec:conclusions}.

\section{Method and Model Parameters}\label{sec:method}

We start by reviewing the physics of the Ly-$\alpha$ damping wing, which is the key feature for probing the IGM neutral fraction with GRB afterglows \citep{MiraldaEscude:1997qb}. The Ly-$\alpha$ optical depth close to line center is large even in the case of a highly ionized universe at the redshifts of interest, as evident in Eq.~\ref{eq:taugp} below.  Consequently, the traditional Ly-$\alpha$ forest blueward of Ly-$\alpha$ at the source redshift is expected to be highly opaque at $z \gtrsim 6$ or so (as also indicated by current observations, e.g., \citealt{Fan:2005es,Yang2020}), regardless of the precise ionization state of the IGM. If significantly neutral gas -- with a neutral hydrogen fraction of order unity -- remains in the IGM, however, the broad damping wing of the Ly-$\alpha$ line owing to its natural line width, leads to an extended absorption feature redward of Ly-$\alpha$ at the source redshift. In contrast, the probability of absorption in the wing is negligible if the IGM is instead highly ionized.  

In what follows, we ignore the impact of peculiar velocities, consider only the wing of the Ly-$\alpha$ line (i.e., we do not treat absorption in the Doppler core), and approximate the gas density by its cosmic mean value. These should be good approximations for our goal of modeling the extended damping wing redward of line center. Further, we adopt the approximation to the line profile in \citet{MiraldaEscude:1997qb}, which includes classical Rayleigh scattering and the Lorentzian form for the natural line width (see \citealt{Lee2013} for refinements, which we will neglect here as they should be unimportant for our forecasts). 

In this case, consider a significantly neutral portion of the IGM with volume-averaged neutral fraction, $\avg{x_{\rm{HI}}}$, between redshift $z_{\mathrm{beg}}$ and $z_{\mathrm{end}}$ along the line of sight to a GRB afterglow at some higher source redshift, $z_{\rm{s}}$. For simplicity, in our modeling we place each GRB at the center of its redshift bin and consider bins of width $\Delta z =1$, as described further below.
The optical depth in the damping wing of the Ly-$\alpha$ line from this neutral patch -- at wavelength offset $\Delta \lambda$ from Ly-$\alpha$ at the source redshift -- is denoted by $\tau_{\mathrm{DW}}(\Delta \lambda)$.
Hence the observed wavelength is $\lambda_{\mathrm{obs}} = \lambda_\alpha (1 + z_\mathrm{s}) + \Delta \lambda$, where $\lambda_\alpha=1215.67 \Ang$ is the rest-frame wavelength of the 
Ly-$\alpha$ transition. We further denote the fractional wavelength offset as $\delta = \Delta \lambda/[\lambda_\alpha(1+z_\mathrm{s})]$. The damping wing optical depth is then (see \citealt{MiraldaEscude:1997qb}; we mostly follow the notation in that work):
\begin{align}\label{eq:taudw}
\tau_{\mathrm{DW}}(\Delta \lambda) = & \frac{\tau_{\mathrm{GP}}(z_\mathrm{s})}{\pi} R_\alpha \avg{x_{\rm{HI}}} \left(1 + \delta\right)^{3/2} \nonumber \\  & \times \left[I(x_{\mathrm{beg}}) - I(x_{\mathrm{end}})\right].
\end{align}
Here $R_\alpha$ is a dimensionless constant related to the natural line-width of the Ly-$\alpha$ line, with $R_\alpha = \Lambda_\alpha/(4 \pi \nu_\alpha)$, where $\Lambda_\alpha$ is the Ly-$\alpha$ decay constant and $\nu_\alpha$ is the rest-frame frequency of the Ly-$\alpha$ transition. Numerically, $R_\alpha = 2.02 \times 10^{-8}$. 
Here $\tau_{\mathrm{GP}}$ denotes the Gunn-Peterson optical depth of fully neutral gas at the cosmic mean density \citep{Gunn1965}. The Gunn-Peterson optical depth is:
\begin{align}\label{eq:taugp}
\tau_{\mathrm{GP}} =  4.0 \times 10^5 & \left[\frac{1+z_{\mathrm{s}}}{7}\right]^{3/2} \left[\frac{\Omega_b h^2}{0.0225}\right] \left[\frac{\Omega_m h^2}{0.132}\right]^{-1/2} \nonumber \\ 
&  \times \left[\frac{1-Y}{0.76}\right],
\end{align}
where $\Omega_m$ and $\Omega_b$ are the usual cosmological matter and baryon density parameters, $h$ in the Hubble parameter in units of 100 km/s/Mpc, and $1-Y$ is the hydrogenic mass fraction.
Finally, $x_{\mathrm{beg}} = (1 + z_{\mathrm{beg}})/[(1+z_{\mathrm{s}}) (1 + \delta)]$, $x_{\mathrm{end}} = (1 + z_{\mathrm{end}})/[(1+z_{\mathrm{s}}) (1 + \delta)]$, and $I(x)$ is an integral with
\beq\label{eq:idef}
I(x_{\mathrm{beg}}) - I(x_{\mathrm{end}}) = \int_{x_{\mathrm{end}}}^{x_{\mathrm{beg}}} dx \frac{x^{9/2}}{(1-x)^2}.
\eeq

In what follows, we imagine that each GRB host is located within an ionized region and so $z_{\mathrm{beg}}$ is set by the distance between the GRB host halo and the first significantly neutral region encountered along the line of sight to the observer. As discussed further below, this distance is drawn from a distribution based on numerical simulations of reionization. Next, we approximate the exterior neutral fraction as uniform outside of the ionized bubble around the GRB host. Our baseline model further adopts the average neutral fraction $\avg{x_{\rm HI}}$ at the center of the redshift bin in question and includes contributions only from neutral gas within the redshift bin. In other words, $z_{\mathrm{end}}$ in Eqs.~\ref{eq:taudw} and \ref{eq:idef} is set by the lower redshift end of the redshift bin under consideration. Put differently, we neglect redshift evolution in the neutral fraction across each redshift bin, and ignore contributions from neutral gas at lower redshifts outside of the bin. Although the precise assumptions here are arbitrary and we will test their impact in what follows, note that the damping wing is fairly insensitive to neutral gas at redshifts much smaller than the source redshift (see Appendix B and Figure~\ref{fig:zend_dep}). 

Throughout we assume perfect knowledge of the GRB host redshift. We suppose that the GRB redshifts may be determined accurately through measuring the redshifts of host-galaxy metal lines detected in absorption, perhaps associated with a local DLA (e.g., \citealt{Totani:2005ng}). Alternatively, follow-up observations after the afterglow fades can be used to determine the redshift of the host galaxy using emission lines (e.g., \citealt{Kruhler_2015}). 

\subsection{Assumed Redshift Distribution}\label{sec:GRB_zdist}

\begin{figure}[htpb]
\bc
\includegraphics[width=1.0\columnwidth]{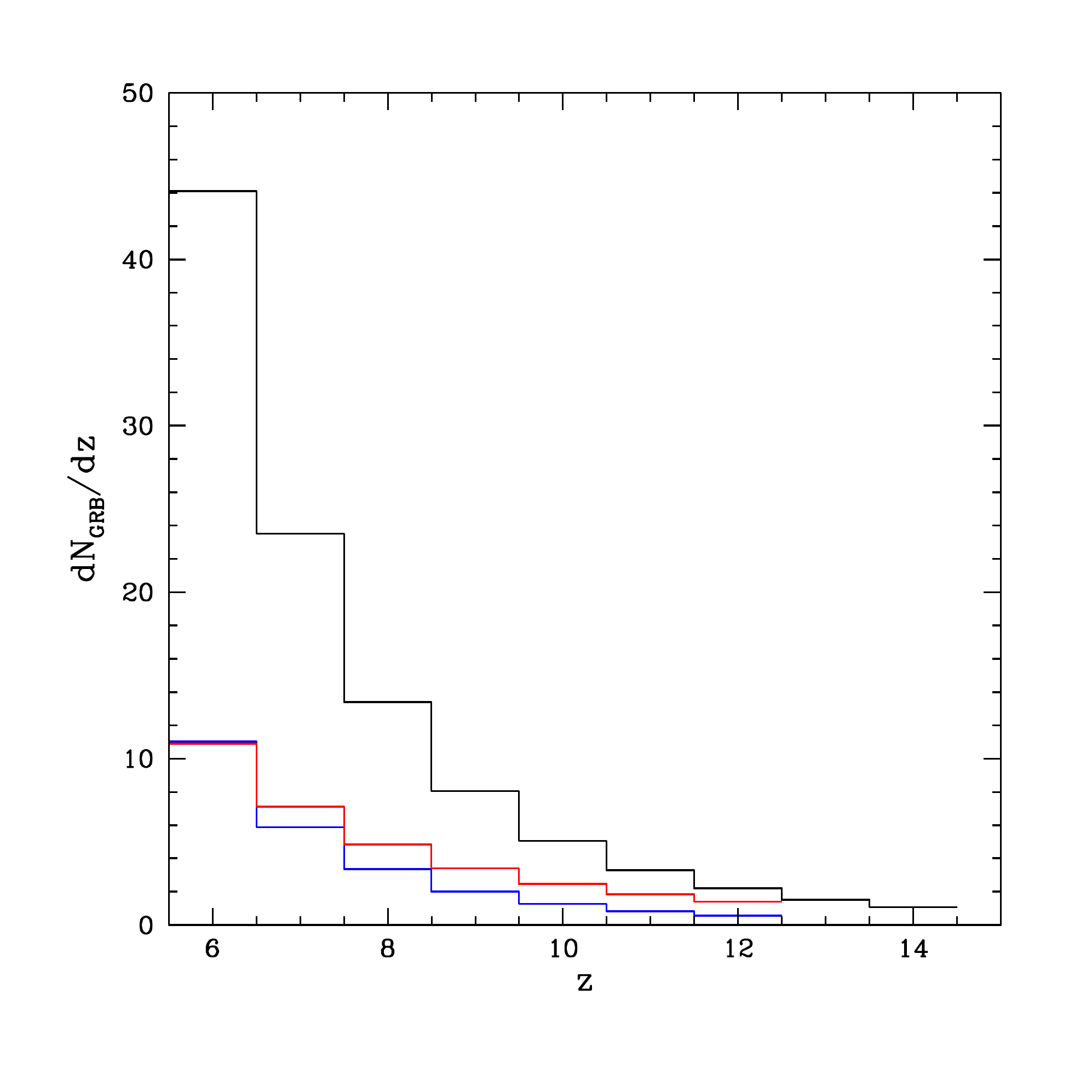}
\caption{Assumed redshift distributions for the rate of long-duration GRBs during the EoR. The blue histogram shows our fiducial model in which the GRB rate traces the SFRD, and the total number
of GRBs above $z=6$ is normalized to 20, as forecast for the Gamow Explorer. The red histogram shows an alternate case, in which the GRB rate is enhanced at high redshift (see text), which is also considered in what follows. Finally, the black histogram shows the upper end of the predicted GRB rate for the planned Thesesus mission.}
\label{fig:grb_histos}
\ec
\end{figure}

The first key ingredient in our forecasts is the assumed redshift distribution of GRB afterglows. 
Provided the GRB rate traces the SFRD, the number of GRBs per unit redshift may be written as \citep{Kistler:2007ud}:
\beq\label{eq:dndz}
\frac{dN_{\mathrm{GRB}}}{dz} = N_0 \frac{\dot{\rho}_\star}{1 + z} \frac{d V_{\mathrm{co}}}{dz},
\eeq
where $\dot{\rho}_\star$ denotes the SFRD, i.e.,  the star formation rate per co-moving volume per unit time, $d V_{\mathrm{co}}/dz$ is the co-moving volume per unit redshift, and $N_0$ is a normalization constant (with units of observing time). The factor of $1 + z$ in the denominator accounts for time dilation. We adopt the empirical fitting formula for the SFRD from \cite{Madau:2014bja}. We ignore any redshift dependence in the ability of the upcoming surveys to detect the GRBs and determine their redshifts in this equation, which should be a good approximation for the bright bursts targeted by the Gamow Explorer and related missions.

Our fiducial model fixes the normalization constant by requiring $N_{\mathrm{GRB}}(\geq z=6) = \int_6^\infty dz\, dN_{\mathrm{GRB}}/dz = 20$. 
This is based on detailed projections of the number of high redshift GRBs detectable with the Gamow Explorer \citep{White2020}, including models for the GRB luminosity function, prompt emission and afterglow spectral shape, observing time, and the Gamow instrumental sensitivity \citep{Ghirlanda2015}. In order to explore some of the uncertainties involved we also consider a model with an enhanced GRB rate at higher redshifts: in this alternate scenario, we multiply the left-hand side of Eq~\ref{eq:dndz} by $(1+z)^{1.5}$, normalized to the same $dN_{\mathrm{GRB}}/dz$ at $z=6$. This case yields $31$ GRBs at $z \geq 6$. This is intended to roughly represent scenarios where either the SFRD estimates from \cite{Madau:2014bja} are incomplete at $z \gtrsim 6$, owing for example to star formation in faint galaxies, or possible evolution in the GRB to star-formation rate relationship. 
Finally, in order to explore the longer term best-case scenario prospects we also consider a case with $N_{\mathrm{GRB}}(z \geq 6)=80$ GRBs. This is at the upper end of the range of 30-80 $z \geq 6$ GRBs forecast for the Theseus mission \citep{Amati:2017npy}.

The GRB redshift distributions in each of these three models are shown in Figure~\ref{fig:grb_histos}, for redshift bins of width $\Delta z=1$. In what follows, we adopt the nearest integer to the average afterglow count forecast in each bin.
In the blue and red historgram models, $11$ afterglows lie in the redshift bin centered around $z_{\rm s}=6$, and at least one GRB is detectable out to the $z_{\rm s}=10$ bin in the fiducial case and $z_{\rm s}=12$ in the  more optimistic scenario. In our fiducial reionization history, described below, these redshifts span most of the reionization process (see Figure~\ref{fig:xhi_vs_z}), and should push past the highest redshift that will be probed using quasars, even with futuristic surveys, as mentioned in the Introduction). In the longer term, the optimistic upper-end of the Theseus projections gives $44$ GRBs in our $z_{\rm s}=6$ redshift bin, $5$ in the $z_{\rm s}=10$ bin and one GRB, on average, in the $z_{\rm s}=14$ bin.\footnote{Note that the total number of GRBs in our histogram exceeds 80 for Theseus because the lower-end of our redshift $z_{\rm s}=6$ bin extends to $z_{\rm s}=5.5$. In our fiducial model, described in the next section, reionization completes at redshift slightly below $z=6$ as may, in fact, be suggested by observations, e.g., \cite{Lidz:2007mz,Mesinger2010,Malloy:2014tba,Kulkarni:2018erh,Keating:2019qda}.} 

\subsection{Fiducial reionization history}\label{sec:fiducial_history}

\begin{figure}[htpb]
\bc
\includegraphics[width=1.0\columnwidth]{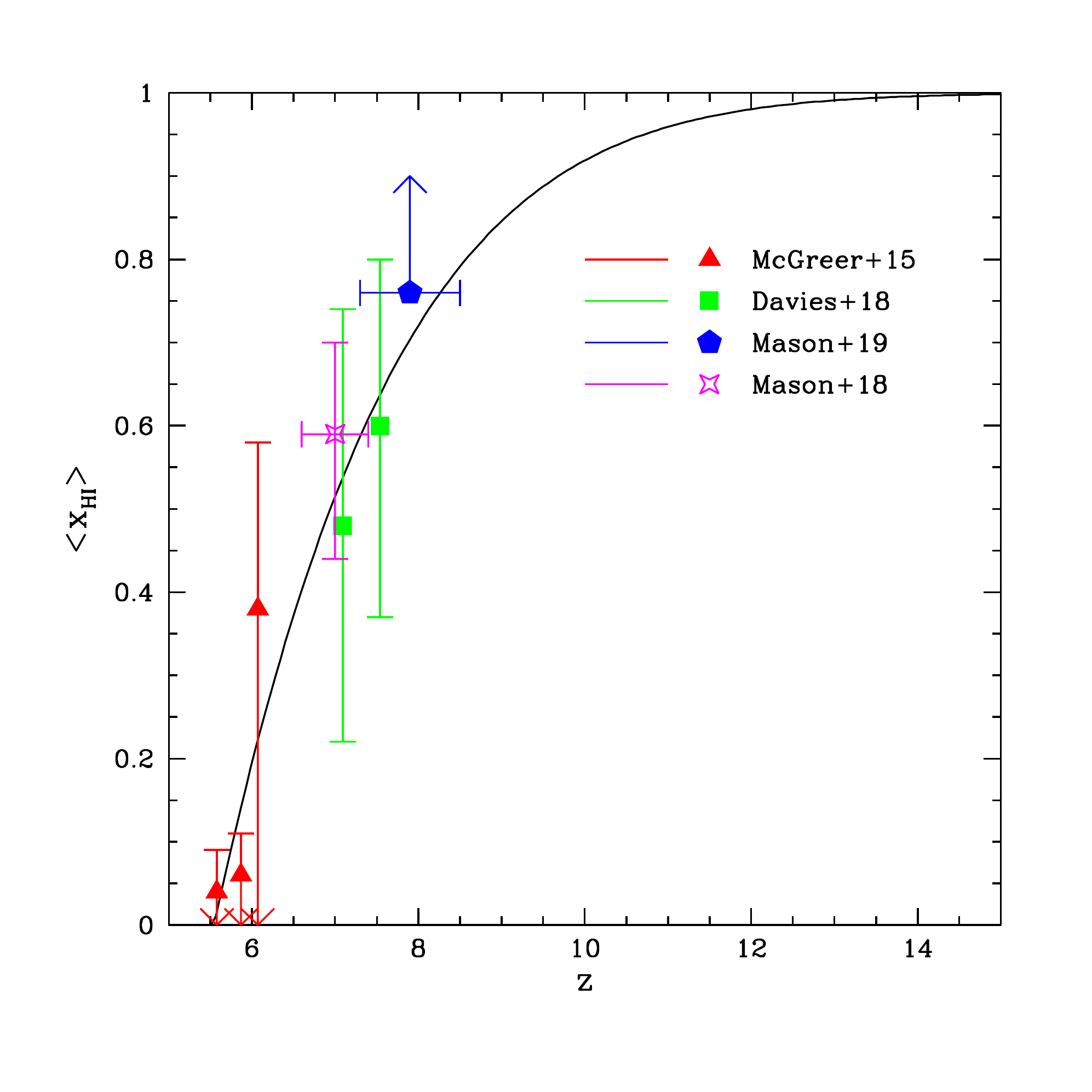}
\caption{Fiducial model reionization history compared to some current observational constraints. The solid black curve shows our fiducial theoretical model described in the text below. The red triangles show upper limits on the neutral fraction from the dark pixel fraction in the Ly-$\alpha$ and Ly-$\beta$ forests \citep{McGreer:2014qwa}. The green squares show estimates of the neutral fraction from damping wing measurements towards quasars \citep{Davies:2018yfp}. Finally, the blue pentagon and magenta star, respectively, show a lower limit and a measurement of the neutral fraction using observations of Ly-$\alpha$ emission lines towards Lyman-break galaxies \citep{Mason:2017eqr,Mason:2019ixe}. In all cases, the error bars are $1-\sigma$ confidence estimates.}
\label{fig:xhi_vs_z}
\ec
\end{figure}

Our fiducial reionization history model is based on an approximate photon-counting description of reionization, in which the volume-averaged ionization fraction is
determined through the following equation \citep{Shapiro87,Madau:1998cd}:
\beq\label{eq:xofz}
\frac{d \avg{x_i}}{dt} = \zeta \frac{d f_{\mathrm coll}(>M_{\mathrm{min}})}{dt} - \frac{\avg{x_i}}{\bar{t}_{\mathrm{rec}}}.
\eeq
Here the first term describes the rate of ionizing photon production, which is assumed to be proportional to the time derivative of the halo collapse fraction, above a minimum galaxy host halo mass, $M_{\mathrm{min}}$. The second term accounts for recombinations, and $\bar{t}_{\mathrm{rec}}$ gives the average time for ionized gas to recombine, which may be written as $\bar{t}_{\mathrm{rec}}= C/[\alpha_B(T) \avg{n_e(z)}]$ where C is a clumping factor, $\alpha_B$ is the case-B recombination coefficient of hydrogen, and $\avg{n_e(z)}$ is the proper number density of free electrons at redshift $z$. Our fiducial model adopts an ionizing efficiency parameter, $\zeta=20$, a minimum galaxy host halo mass of $M_{\mathrm{min}}=10^9 M_\odot$, a gas temperature of $T=2 \times 10^4$ K, and a clumping factor of $C=3$ (see e.g., \citealt{Lidz:2015ewe} and references therein for a discussion of these parameters). The halo collapse fraction is calculated using the Sheth-Tormen halo mass function \citep{Sheth2001}. The ionization fraction is fixed to unity below the redshift where the solution of Eq.~\ref{eq:xofz} first reaches this limiting value. The volume-averaged neutral fraction is determined through 
$\avg{x_{\rm HI}(z)} = 1 - \avg{x_i(z)}$. 

The resulting ionization history is shown in Figure~\ref{fig:xhi_vs_z}. We emphasize that the fiducial model here is intended only as a plausible baseline for our ensuing forecasts, and has not been designed or tuned to match current constraints in detail. It is nevertheless in good agreement with current measurements, as illustrated in the figure. Furthermore, the electron scattering optical depth in our fiducial model is $\tau_e=0.51$ (assuming helium is singly ionized along with hydrogen and neglecting the likely small contribution from the second ionization of helium), consistent with recent Planck 2018 measurements: the fiducial model $\tau_e$ value differs from their  TT, TE, EE, lowE, lensing + BAO constraint by less than $1-\sigma$ \citep{Aghanim:2018eyx}.

\subsection{Bubble size distribution}\label{sec:bubble_dist}

\begin{figure}[htpb]
\bc
\includegraphics[width=1.0\columnwidth]{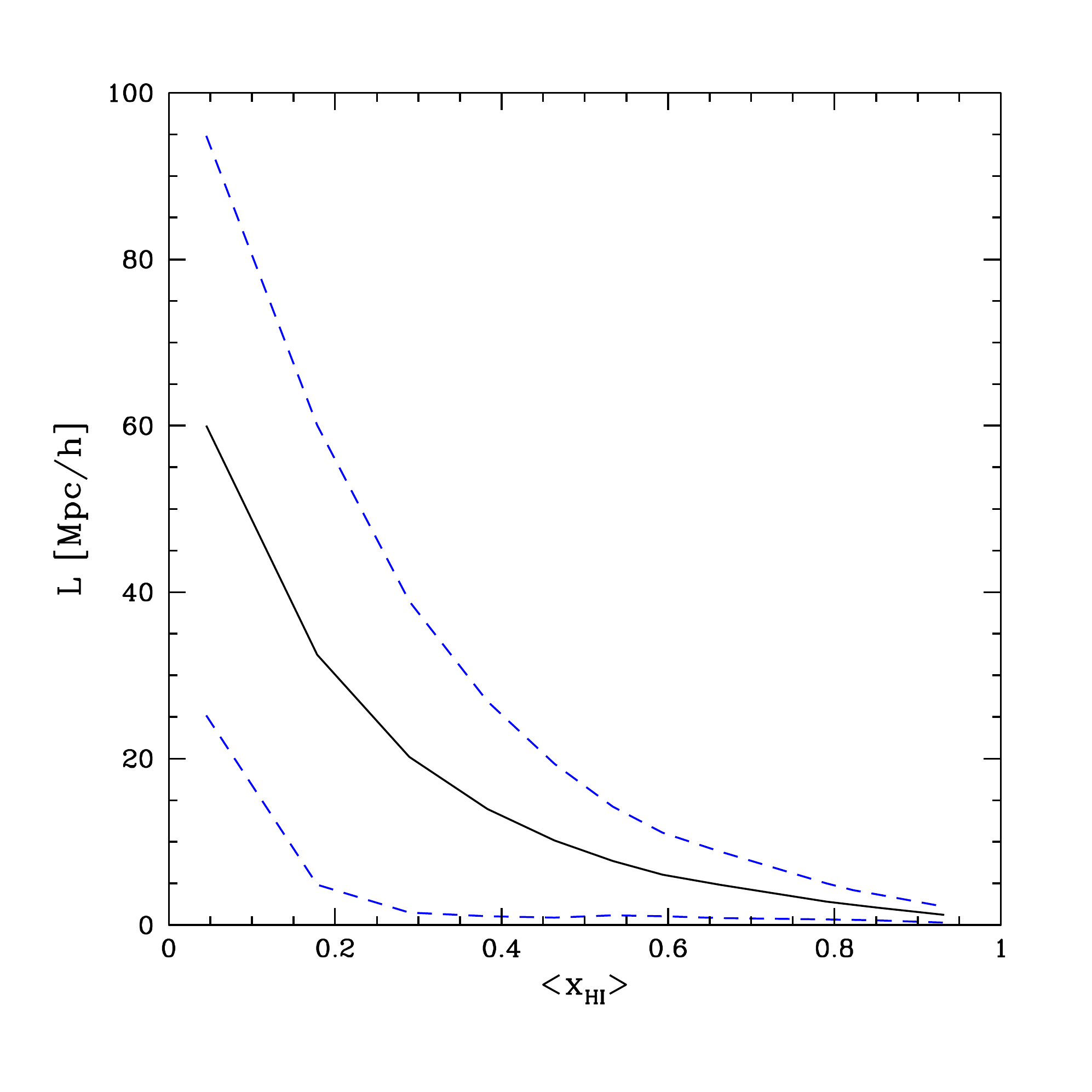}
\caption{The mean bubble (or ionized skewer) size and its scatter -- as a function of the volume-averaged neutral fraction -- extracted from reionization simulations. The black solid line gives the average co-moving distance between a GRB host and the first neutral region, while the blue dashed lines indicate the sightline to sightline scatter around the mean. The mean and scatter grow sharply as reionization proceeds and the average neutral fraction drops.}
\label{fig:bubble_size}
\ec
\end{figure}

The next important ingredient in our model is to account for the ionized bubbles around GRB host halos. In general, these regions reflect the combined ionizing output of many surrounding galaxies, with the typical size scale growing as ionized bubbles expand and merge over the course of the reionization process (e.g., \citealt{McQuinn:2006et}). 
In order to appreciate the role of the ionized bubbles here, consider 
the example of a GRB that occurs in a large ionized region. More precisely, the relevant thing here is the the extent of the ionized skewer traversed along the line of sight from the GRB to the observer before encountering neutral gas in the IGM.\footnote{We will use the terms ``ionized bubble size'' and ``ionized skewer size'' interchangeably, although ``skewer'' is a more appropriate term for the present application given that GRB afterglows provide a 1-d probe through the IGM.} 
In this case, photons initially just redward of Ly-$\alpha$ at the source redshift, for instance, will redshift significantly before encountering neutral gas and the source will show only weak damping-wing absorption. In contrast, GRB afterglow photons originating in small bubbles or at the edge of larger ones will redshift less before running into neutral gas and the damping wing feature will be stronger.  Therefore the size of the ionized bubbles around GRB host halos plays an important role in determining the strength of the damping wing absorption.

In order to account for this in our model, we turn to the reionization simulations of \cite{McQuinn:2007gm,McQuinn:2007dy}, which treat the radiative transfer of ionizing photons in a post-processing step, on top of an evolved N-body simulation of cosmological structure formation. From a number of simulation snapshots spanning much of the EoR, we extract sightlines in random directions towards dark matter halos with mass between $10^{10} M_\odot \leq M \leq 10^{11} M_\odot$ and record the co-moving distance between each simulated dark matter halo and the first significantly neutral cell (defined by when the cell-averaged neutral fraction exceeds 0.9).\footnote{\cite{McQuinn:2007gm} considered the same basic distribution, but chose a different criterion for the first neutral cell and host halo mass range. Their results are hence slightly different than those here.}
These halos are well-resolved in the simulations and are plausible hosts for GRBs during the EoR. The reionization history in the simulations differs a little from our fiducial model, and so we record the first-crossing distance as a function of the volume-averaged neutral fraction rather than redshift. In each GRB redshift bin, we then determine the mean ionized skewer size and the (1-$\sigma$) scatter from the simulation output with average neutral fraction closest to the fiducial model value at the desired redshift. Relying on simulation outputs at the correct neutral fraction, but slightly different redshifts, should be a good approximation: previous studies have shown that the spatial structure of reionization is sensitive mainly to the average ionization fraction and is fairly independent of the precise redshift when a given stage of reionization is reached \citep{McQuinn:2006et}. 
This first-crossing distance then sets the upper limit of the integral over redshift, $z_{\mathrm{beg}}$ in Eqs~\ref{eq:taudw} and \ref{eq:idef}.\footnote{Towards the end of reionization in our model, the co-moving distance to the first neutral region, which sets $z_{\mathrm{beg}}$, may occasionally exceed the distance from bin center to the lower-redshift edge of a bin, which sets $z_{\mathrm{end}}$. In this case, the line of sight is considered to traverse entirely ionized gas and leads to negligibly small damping wing absorption in the model.} 

To briefly reiterate, the main approximation in our scheme is to ignore spatial variations in the neutral fraction outside of the first ionized bubble. We expect this to be most accurate in the early stages of reionization when the neutral fraction is large and the size of typical ionized bubbles is small relative to the extent of the damping wing. Especially towards the end of reionization, a more elaborate parameterized model involving the size of the first neutral region, the extent of the second ionized bubble encountered, etc. may be required for more accurate results. 
We confine our current study, however, to the simpler three-parameter model.

Figure~\ref{fig:bubble_size} shows the resulting distribution of ionized skewer sizes. The mean skewer size (in co-moving units) evolves very strongly during the EoR, ranging from roughly $\sim 1$ Mpc/$h$ when the universe is $\sim 95\%$ neutral, to $\sim 10$ Mpc/$h$ when it is $\sim 50\%$ neutral, and to $\sim 60$ Mpc/$h$ when about $\sim 5\%$ of the IGM volume is neutral. The scatter also grows strongly as reionization proceeds. 

The size distribution of ionized skewers may be conveniently approximated by a lognormal distribution. Specifically, 
\beq\label{eq:longnorm}
\frac{dP_L}{dL} = \frac{1}{\sqrt{2 \pi \tilde{\sigma}^2}}\frac{1}{L} \rm{exp}\left[-\frac{\left(\rm{ln} L - \mu\right)^2}{2 \tilde{\sigma}^2}\right]
\eeq
where $L$ is the bubble size in units of Mpc/$h$. 
The lognormal parameters $\mu$ and $\tilde{\sigma}$ are related to the mean ionized skewer size, $\avg{L}$, and sightline to sightline scatter, $\sigma_L$, via:
\beq\label{eq:mean_bubble}
\mu = \rm{ln} \left[\frac{\avg{L}^2}{\left(\sigma_L^2 + \avg{L}^2\right)^{1/2}}\right],
\eeq
and
\beq\label{eq:sigma_bubble}
\tilde{\sigma}^2 = \rm{ln} \left[\left(\frac{\sigma_L}{\avg{L}}\right)^2 + 1 \right]
\eeq

\subsection{Contamination from Host DLAs}\label{sec:dlas}

In addition to absorption from neutral hydrogen in the IGM, we must account for absorption from neutral hydrogen in the host galaxy. Empirically, as discussed further below, many GRB sightlines show strong DLA absorption. 
We assume that the DLA lies precisely at the redshift of the GRB host galaxy, which is likely an excellent approximation given that long duration GRBs are expected to trace star-forming regions, and most star formation occurs in relatively small mass halos at the redshifts of interest. For instance, the circular velocity of a $10^{10} M_\odot$ halo at $z_{\rm s}=8$ is $v_{\rm circ}=74$ km/s \citep{Barkana2001} and so a DLA moving at the halo circular velocity relative to the host galaxy is offset in wavelength by only $\Delta \lambda/\lambda \sim v_{\rm circ}/c \sim 2 \times 10^{-4}$, comparable to the spectral resolution assumed in what follows (\S \ref{sec:fisher}). This is hence a negligible effect, although a DLA in a neighboring halo could in principle be more of a concern; an unlikely coincidence would, however, be required for the neighboring absorber to produce a large column density.

Since we are interested in the absorption in the wing of the line, we approximate the optical depth of the DLA at wavelength offset $\Delta \lambda$ using a Lorentzian profile:
\beq\label{eq:tau_dla}
\tau_{\mathrm{DLA}}(\Delta \lambda) = N_{\rm{HI}} \frac{\sigma_\alpha}{\pi} \frac{R_\alpha}{(\Delta \lambda/\lambda)^2 + R_\alpha^2},
\eeq
where $\sigma_\alpha = 3 \lambda_\alpha^2 \Lambda_\alpha/(8 \pi \nu_\alpha) = 4.48 \times 10^{-18}$ cm$^{2}$ is the frequency-integrated cross section. The transmission through the IGM accounting for the IGM damping wing and the DLA is then given from Eqs \ref{eq:taudw} and \ref{eq:tau_dla} as:
\beq\label{eq:transmission}
f = e^{-\tau} = e^{-(\tau_{\mathrm{DW}} + \tau_{\mathrm{DLA}})} .
\eeq

\begin{figure}[htpb]
\bc
\includegraphics[width=1.0\columnwidth]{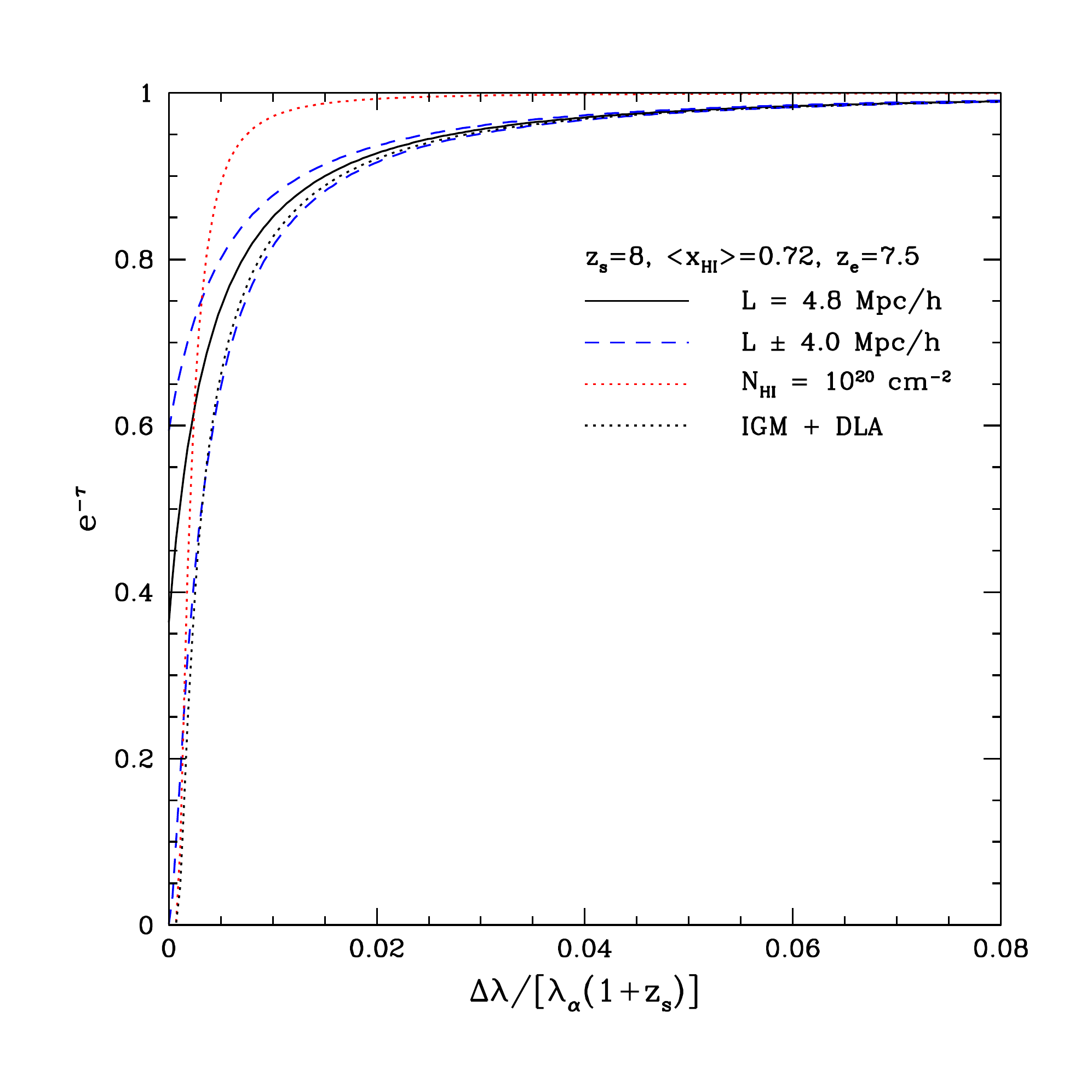}
\caption{Example damping wing profile. The black solid line shows the IGM damping wing in our fiducial model at $z_{\rm s}=8$ for the mean ionized skewer size, while the blue dashed lines show the profiles for skewers of length $\pm 1-\sigma$ above and below the mean. In each case $z_{\rm end} = 7.5$ (see text). The red dotted line shows the profile of a DLA at the host redshift with a column density of $N_{\rm{HI}} = 10^{20}$ cm$^{-2}$. The black dotted line shows the transmission including both the DLA optical depth and the IGM contribution for the mean skewer size. 
The more gradual IGM profile is distinguishable from the DLA profile for the parameters shown, although the spatial variations in the ionized skewer size lead to fairly strong variations in the transmission close to line center. 
}
\label{fig:examp_dwing}
\ec
\end{figure}

We will explore the redshift and parameter dependence of the transmission profiles in more detail in the next section, but Figure~\ref{fig:examp_dwing} gives an illustrative example for starters. Here we consider a GRB source at $z_{\mathrm{s}} = 8$, in which case the IGM is 70\% neutral in our fiducial model (Figure~\ref{fig:xhi_vs_z}). The mean size of ionized skewers at this neutral fraction is $\avg{L} = 4.8$ Mpc/$h$, while the sightline to sightline dispersion is $\sigma_L = 4.0$ Mpc/$h$ (Figure~\ref{fig:bubble_size}). In this example, the figure illustrates that the DLA dominates the absorption close to line center for the average ionized skewer size, but ionized skewers with size $1-\sigma$ lower than the mean value produce comparable absorption at line center to the DLA. However, for any skewer size within the $\pm 1-\sigma$ range around the mean in this example, we expect the IGM contribution to the damping wing to dominate over the DLA component far from line center (for $N_{\rm HI} = 10^{20}$ cm$^{-2}$). That is, the IGM component, scaling roughly as $\tau_{\rm DW} \propto 1/(\Delta \lambda)$ (Eq.~\ref{eq:taudw})\footnote{This follows from the small $\delta$ limit of Eq.~\ref{eq:taudw} provided $z_{\mathrm{end}}$ and $z_{\mathrm{beg}}$ are close to $z_{\mathrm{s}}$.}, is more extended than the DLA contribution which goes like $\tau_{\rm DLA} \propto 1/(\Delta \lambda)^2$ (Eq.~\ref{eq:tau_dla}). As fleshed out further in \S \ref{sec:fisher} below, larger column density DLAs can nevertheless swamp the IGM damping wing contribution, especially at lower redshifts where the neutral fraction is smaller, the universe is less dense, and the ionized skewers from the GRB host halo tend to be larger.

\begin{figure}[htpb]
\bc
\includegraphics[width=1.0\columnwidth]{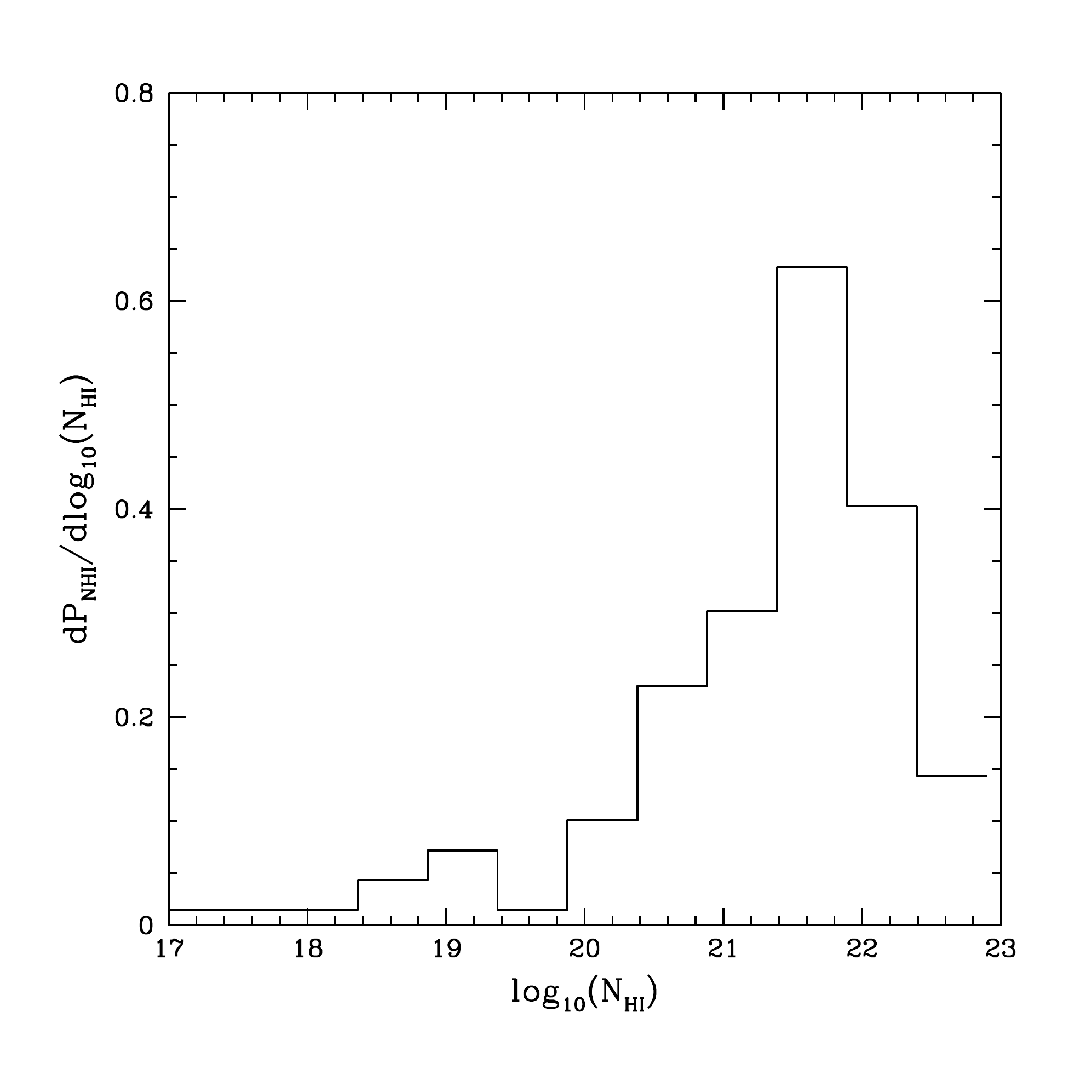}
\caption{The probability distribution of DLA column density from the current sample of 138 GRB afterglows at $1.6 < z < 6.7$. This is based on the data tabulated in \cite{Tanvir:2018pbq}.}
\label{fig:pdf_nhi}
\ec
\end{figure}

A key, yet uncertain, issue then regards the probability distribution of DLA column densities from local host absorption around GRBs during the EoR. The largest empirical sample currently addressing this question is published in \cite{Tanvir:2018pbq}; these authors determined the column density distribution from a sample of 138 GRB afterglows ranging in redshift from $1.6 < z < 6.7$. The probability distribution from the tabulated data in this study is shown in Figure~\ref{fig:pdf_nhi}. Interestingly, \cite{Tanvir:2018pbq} derive an average ionizing photon escape fraction of only $\avg{f_{\rm esc}} = 0.005$, with a 98\% upper limit of $\avg{f_{\rm esc}} \leq 0.015$. Indeed, the median column density in this sample is $N_{\rm HI}=10^{21.59}$ cm$^{-2}$, relatively few sightlines have $N_{\rm HI} \leq 10^{20}$ cm$^{-2}$, and only a tiny fraction of sightlines have column densities low enough to allow appreciable ionizing photon escape fractions. 

An important question then is how strongly the probability distribution of DLA column densities evolves with redshift. The current sample includes only $7$ lines of sight above $z > 5$, and while there is currently no evidence for redshift evolution in the column density probability distribution function (PDF) \citep{Tanvir:2018pbq}, a larger data set will be required to test this more sharply. In fact, provided this distribution is indeed a reliable indicator of the escape fraction of ionizing photons, one can indirectly argue that the distribution {\em must evolve strongly}: otherwise it is hard to understand how the universe could be reionized by $z \lesssim 5.5$ (as is observed) given the escape fraction estimate in \citet{Tanvir:2018pbq}. For instance, escape fractions larger than roughly $f_{\rm esc} \gtrsim 0.1$ are required to reionize the universe with star-forming galaxies, subject somewhat to uncertainties in: faint-end extrapolations down the galaxy UV luminosity function, the ionizing spectral shape, and the clumping factor of the IGM (e.g., \citealt{Bouwens2016}). This is a factor of $\sim 20$ larger than the best fit value from the current GRB samples, which are however concentrated at lower redshifts. The strong evolution implied here is at least broadly consistent with inferences of the escape fraction at $z \sim 2.5-5$ based on comparing galaxy UV luminosity functions to the ionizing emissivity extracted using the transmission through the Ly-$\alpha$ forest at these redshifts, along with estimates of the mean free path of ionizing photons \citep{Faucher-Giguere:2019kbp,FaucherGiguere:2008rc}.
Nevertheless, our fiducial model adopts the current DLA column density PDF since it is hard to predict how this may evolve towards higher redshift. We should keep in mind that this may be a pessimistic assumption (see also \S \ref{sec:fisher_cont}).

\section{Fisher Matrix formalism}\label{sec:fisher}

In order to forecast the constraints on the ionization history, we adopt a Fisher matrix approach. The Fisher matrix calculations assume a quadratic expansion around the logarithm of the likelihood function, which in turn peaks for a set of specified fiducial model parameters. As shown in Appendix A, we can incorporate the expected broad distributions in the ionized skewer size and DLA column density PDFs by first computing a conditional Fisher matrix for sightlines with specified bubble size $L$ and column density $N_{\rm{HI}}$. We then find the expected error bar on the neutral fraction for that $L$ and $N_{\rm{HI}}$ by inverting the conditional Fisher matrix. Finally, the error bar on the neutral fraction for an ensemble of GRB afterglow sightlines can be determined by computing the average of the neutral fraction's inverse-variance, weighted by the bubble size and column density PDFs (see Eqs.~\ref{eq:fisher_xx_final} and \ref{eq:error_x_final}). The error bar in each redshift bin is scaled with $1/\sqrt{N_{\rm GRB}(z_{\rm s})}$ where $N_{\rm GRB}(z_{\rm s})$ is the number of independent afterglow spectra expected in the bin centered around source redshift $z_{\rm s}$.
For each sightline, we adopt a parameter vector $\bf{q}$ with three components, $(q_{\rm xHI}, q_L, q_{\rm NHI})$. Here each component of $\bf{q}$ describes the fractional variation around the fiducial model value. For example, $q_{\rm xHI} = (\avg{x_{\rm HI}} - \avg{x_{\rm HI, fid}})/\avg{x_{\rm HI, fid}}$. Note also that the parameters $q_L$ and $q_{\rm NHI}$ vary in meaning depending on the value of $L$ and $N_{\rm{HI}}$ for each sightline. 

The conditional Fisher matrix for a sightline with bubble size $L$ and column density $N_{\rm{HI}}$ may be computed as:
\beq\label{eq:fisher_conditional}
\tilde{F}_{ij}(L,N_{\rm{HI}}) = \sum_k \frac{\partial f(k|L, N_{\rm{HI}})}{\partial q_i}\frac{\partial f(k|L, N_{\rm HI})} {\partial q_j} \frac{1}{\mathrm{var}\left[f(k)\right]}.
\eeq
Here the sum is over spectral pixels, which are taken to be uniformly spaced in $\rm{ln}(\lambda)$ with $\Delta \lambda/\lambda = 1/3,000$, i.e., we consider a spectral resolution
of $R=\lambda/\Delta \lambda = 3,000$. We sum over all pixels out to $1,500 \Ang$ redward (observed frame) of Ly-$\alpha$ at the source redshift. Appendix B demonstrates that our results are insensitive to this choice (see Figure~\ref{fig:almax_dep}).
In general, we assume that the noise in each afterglow spectrum is set by a combination of Poisson fluctuations in the photons from the source itself and
a contribution from the sky background. In this case,
\beq\label{eq:var_flux}
\mathrm{var}\left[f(k)\right] = \frac{N_{\mathrm{c}} f(k) + N_{\mathrm{sky}}}{N_{\mathrm{c}}^2},
\eeq
where $N_{\mathrm{c}}$ is the number of photons per spectral resolution element received from the source at the continuum and $N_{\mathrm{sky}}$ is the same for the sky background. We adopt a flat continuum spectrum, which should be a very good approximation across the range of wavelengths considered in Eq.~\ref{eq:fisher_conditional}.\footnote{Throughout we assume that the unabsorbed continuum level is determined perfectly, i.e.,  we neglect continuum fitting errors in our analysis.} 
The average number of photon counts in spectral pixel $k$ is then given by $N_{\mathrm{c}} f(k)$, i.e., the observed photon count is the number of photons at the continuum multiplied by the transmission. 
The signal-to-noise-ratio squared at the continuum is hence $\mathrm{SNR}^2 = N_{\mathrm{c}}^2/(N_{\mathrm{c}} + N_{\mathrm{sky}})$. 

Our baseline assumption is that the signal-to-noise ratio at the continuum per spectral pixel is $\mathrm{SNR}=20$, and that $N_{\mathrm{sky}} \gg N_{\mathrm{c}}$. That is, we assume that the observations are in the sky background dominated limit, where the source counts make a negligible contribution to the transmission variance.
In this case, $\mathrm{var} \left[f(k)\right] = N_{\mathrm{sky}}/N_{\mathrm{c}}^2 = 1/\mathrm{SNR}^2$. We discuss the possibilities for obtaining our baseline SNR value in \S \ref{sec:snr_dep} and show that the sky background dominated limit applies for ground-based observations, but that space-based observations with the JWST will be in the source dominated limit. We compare the sky background limited and source dominated cases in Appendix B and Figure~\ref{fig:alpha_dep}.  

We consider a redshift independent SNR for the purposes of having a single baseline noise value; this helps us understand the overall requirements for spectroscopic follow-up observations. In practice, the sky background is a strongly increasing function of wavelength and so in practice it will be hard to achieve the requisite SNR from the ground for higher redshift bursts (see \S \ref{sec:snr_dep}). However, in the source-dominated case applicable to JWST space-based observations, a redshift independent SNR may be a decent approximation. This is the case because given afterglows observed at a fixed time interval after the explosion, higher redshift ones are seen at an earlier time in the source frame -- owing to cosmological time dilation -- when they are brighter, compensating for the dimming from the increased luminosity distance (e.g., \citealt{Barkana:2003ja}).
We will consider variations around the SNR assumptions in \S \ref{sec:snr_dep}. 

\subsection{Parameter Derivatives}

\begin{figure}[htpb]
\bc
\includegraphics[width=1.0\columnwidth]{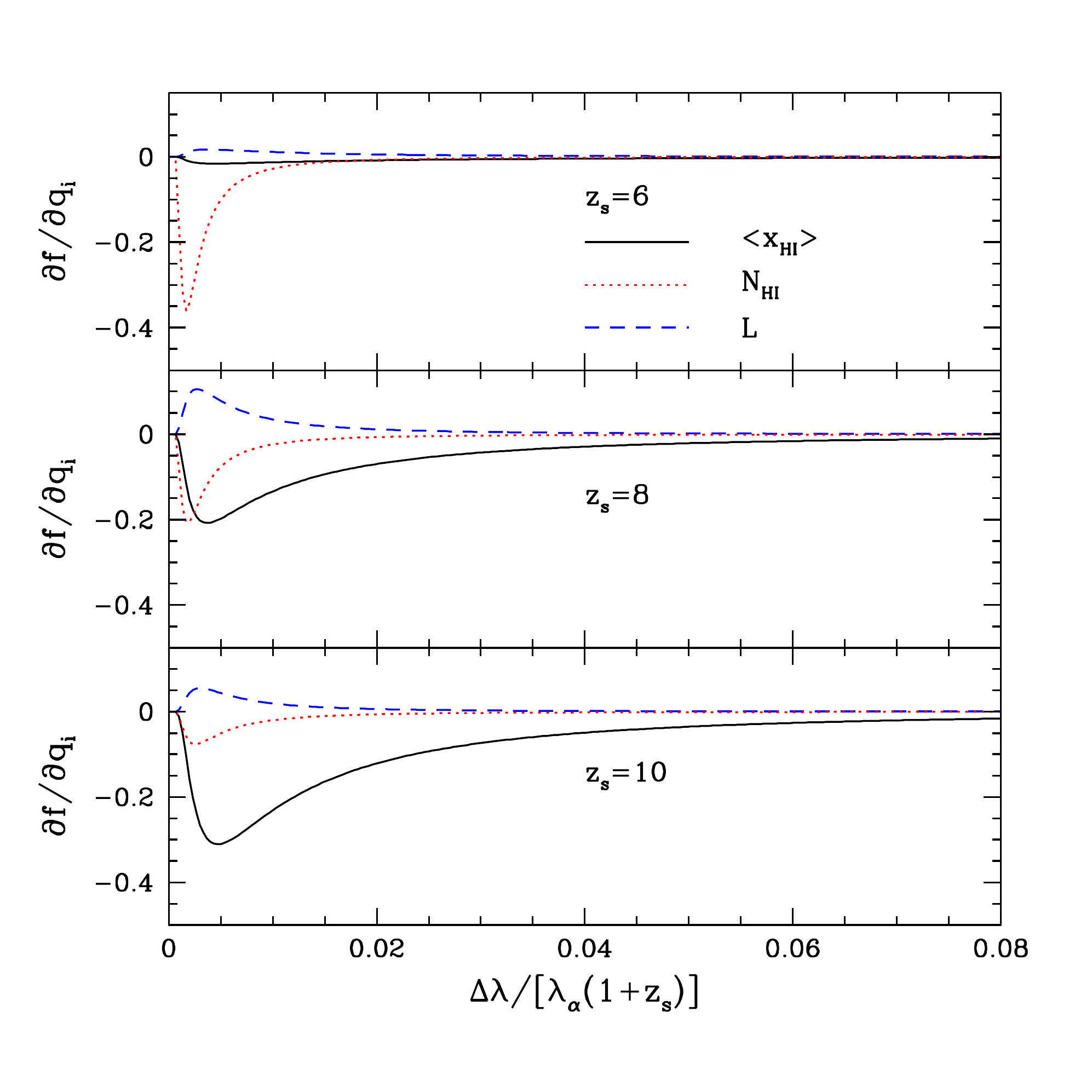}
\caption{Examples of the derivatives which enter our Fisher matrix computations. {\em Top panel}: The transmission derivatives at $z_{\rm s}=6$ for a fiducial model with $\avg{x_{\rm HI}}=0.21$, $L=\avg{L} = 32$ Mpc/$h$, and $N_{\rm HI} = 10^{20}$ cm$^{-2}$. The black solid line shows the derivative with respect to $q_{\rm xHI}$, the red dotted line is the $q_{\rm NHI}$ derivative, and the blue dashed line gives the $q_L$ derivative. {\em Middle panel}: The same, except for $z_{\rm s}=8$, $\avg{x_{\rm HI}} = 0.72$, $L=\avg{L}=4.8$ Mpc/$h$, and  $N_{\rm HI} = 10^{20}$ cm$^{-2}$. {\em Bottom panel}: Identical to the upper two panels, but here we consider $z_{\rm s}=10$, $\avg{x_{\rm HI}} = 0.92$, $L=\avg{L}=1.2$ Mpc/$h$, and  $N_{\rm HI} = 10^{20}$ cm$^{-2}$. }
\label{fig:fisher_derivs}
\ec
\end{figure}

It is instructive to first examine some examples of the transmission derivatives with respect to model parameters, as enter into Eq.~\ref{eq:fisher_conditional}.  Specifically, Figure~\ref{fig:fisher_derivs} shows derivatives with respect to the neutral fraction, bubble size, and DLA column density parameters for the case of our fiducial reionization history with $L=\avg{L}$ and $N_{\rm{HI}}=10^{20}$ cm$^{-2}$ at each of $z_{\rm s}=6, 8$, and $10$ (top, middle, and bottom panels, respectively). In general, increasing the DLA column density leads to reduced transmission relatively close to the redshift of the source, while increasing the bubble size parameter has the opposite effect, as expected. Increasing the neutral fraction reduces the transmission, and the impact of the neutral fraction parameter is more extended in wavelength than either of the DLA or bubble size parameters, owing to the gradual fall-off in the IGM damping wing profile of Eq.~\ref{eq:taudw}.  The fairly similar shape, yet opposite sign, of the derivatives with respect to the bubble size and column density parameters suggests that there will be a fairly strong degeneracy direction between these parameters. That is, the enhanced transmission with increasing bubble size can be compensated by raising the DLA column density. 

Turning to the redshift dependence in Figure~\ref{fig:fisher_derivs}, one can see that the neutral fraction parameter becomes more important than the DLA and bubble size parameters as one moves to higher redshift. This results because the strength of the IGM damping wing grows as the universe becomes more neutral and denser towards higher redshift and as the bubble sizes shrink. Note further that the transmission changes less with increasing DLA column density towards high redshift because the derivatives are taken around a fiducial transmission profile, which is more absorbed at high redshift. For the example DLA column density and bubble sizes, the IGM damping wing signature is swamped at $z_{\rm s} \sim 6$ but becomes more prominent at $z_{\rm s} \gtrsim 8$. On the other hand, we expect to detect more GRB afterglows near $z_{\rm s} \sim 6$ than at $z_{\rm s} \sim 8-10$ (see Figure~\ref{fig:grb_histos}) and so there is a trade-off between the weaker damping wing signatures towards low redshift and the more abundant data samples available there. Also, we should keep in mind that the IGM damping wing signature is more detectable towards sightlines with smaller ionized bubbles and lower column density DLAs (see \S \ref{sec:forecasts}). 

\subsection{Parameter Degeneracies}

\begin{figure}[htpb]
\bc
\includegraphics[width=1.0\columnwidth]{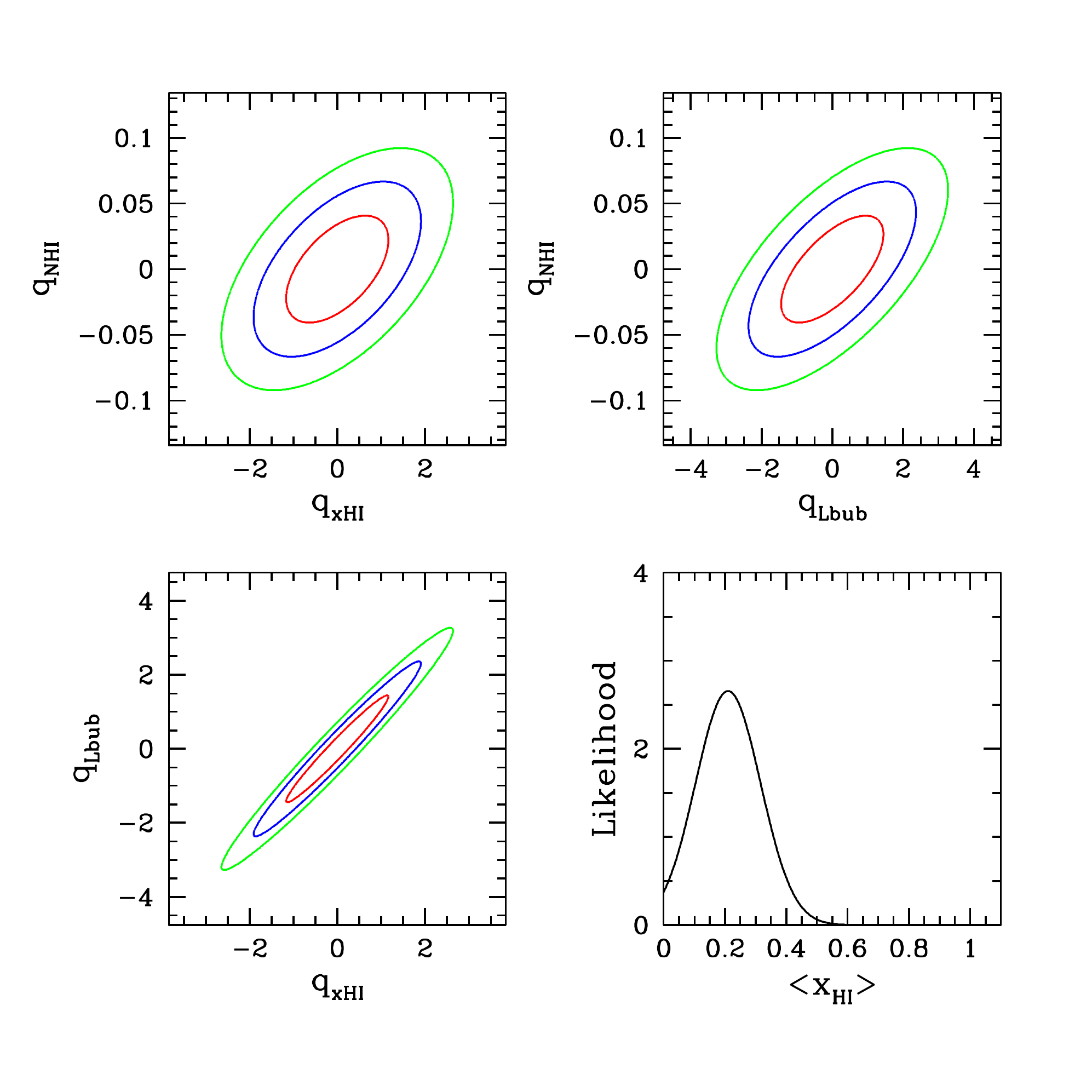}
\caption{Parameter degeneracies for the simplified case that all of the sightlines in the $z_{\rm s}=6$ bins have bubble size $L=\avg{L}$ and $N_{\rm HI} = 10^{20}$ cm$^{-2}$. The ellipses show $1-\sigma$ (red), $2-\sigma$ (blue), and $3-\sigma$ confidence intervals (green). The bottom right-hand corner gives the resulting 1-d likelihood, marginalized over bubble size and DLA column density parameters,  for the average neutral fraction parameter in this redshift bin. We refine this error bar estimate in the next section by accounting for the full distribution of ionized skewer sizes and DLA column densities, but this simplified case nevertheless suffices for understanding the degeneracies involved.}
\label{fig:param_degen_z6}
\ec
\end{figure}

In order to quantify the ionization history constraints in detail, we must account for the full probability distributions in bubble size and DLA column density. First, however, it is helpful to consider parameter degeneracies for particular realizations of the bubble size and DLA column density. That is, we can invert the conditional Fisher matrix of Eq~\ref{eq:fisher_conditional} and inspect the resulting parameter ellipses for particular values of $L$ and $N_{\rm HI}$. For simplicity of illustration, we suppose here that all of the sightlines in the redshift bin of interest have the specified bubble size and DLA column density. Here we take our fiducial GRB redshift distribution with 20 $z \geq 6$ GRBs (Figure~\ref{fig:grb_histos}).

Figure~\ref{fig:param_degen_z6} shows an example case at $z_{\rm s}=6$ with $L=\avg{L} = 32$ Mpc/$h$ and $N_{\rm HI} = 10^{20}$ cm$^{-2}$. The first conclusion one can draw from this figure is that the fractional constraints on the DLA column density in this example are much tighter than those forecast for either the bubble size or the neutral fraction. This results because the derivative with respect to the DLA column density parameter is much stronger than the bubble size and neutral fraction parameter derivatives here, as previously illustrated in the top panel of Figure~\ref{fig:fisher_derivs}. \footnote{Note that we do not incorporate a ``physicality prior'' on any of the parameters, although regions of parameter space with $q_i < -1$ are unphysical. This leads to conservative bounds.} The strongest degeneracy direction here is between ionized skewer size and neutral fraction, since one can trade-off for an increasing neutral fraction by boosting the bubble size. Similarly, the column density and bubble size parameters are positively correlated. Naively, the positive correlation between DLA column density and neutral fraction seen in the upper left hand panel of the figure is surprising, but this results because the bubble size is also an important parameter in the problem: one can compensate for an increase in both the neutral fraction and DLA column density by increasing the bubble size. Indeed, if we fix the bubble size parameter entirely we find the opposite degeneracy direction between $q_{\rm NHI}$ and $q_{\rm xHI}$, as expected. In spite of the weak effect from the IGM damping wing in this case, one would still expect an interesting upper bound on the neutral fraction after accumulating the $11$ GRB afterglow spectra in this redshift bin, as illustrated in the bottom right hand corner of Figure~\ref{fig:param_degen_z6}. This panel shows the 1-d likelihood for $\avg{x_{\rm HI}}$ after marginalizing over the DLA column density and bubble size parameters. This estimate needs to be refined, however, to incorporate the DLA and skewer size distributions.

\begin{figure}[htpb]
\bc
\includegraphics[width=1.0\columnwidth]{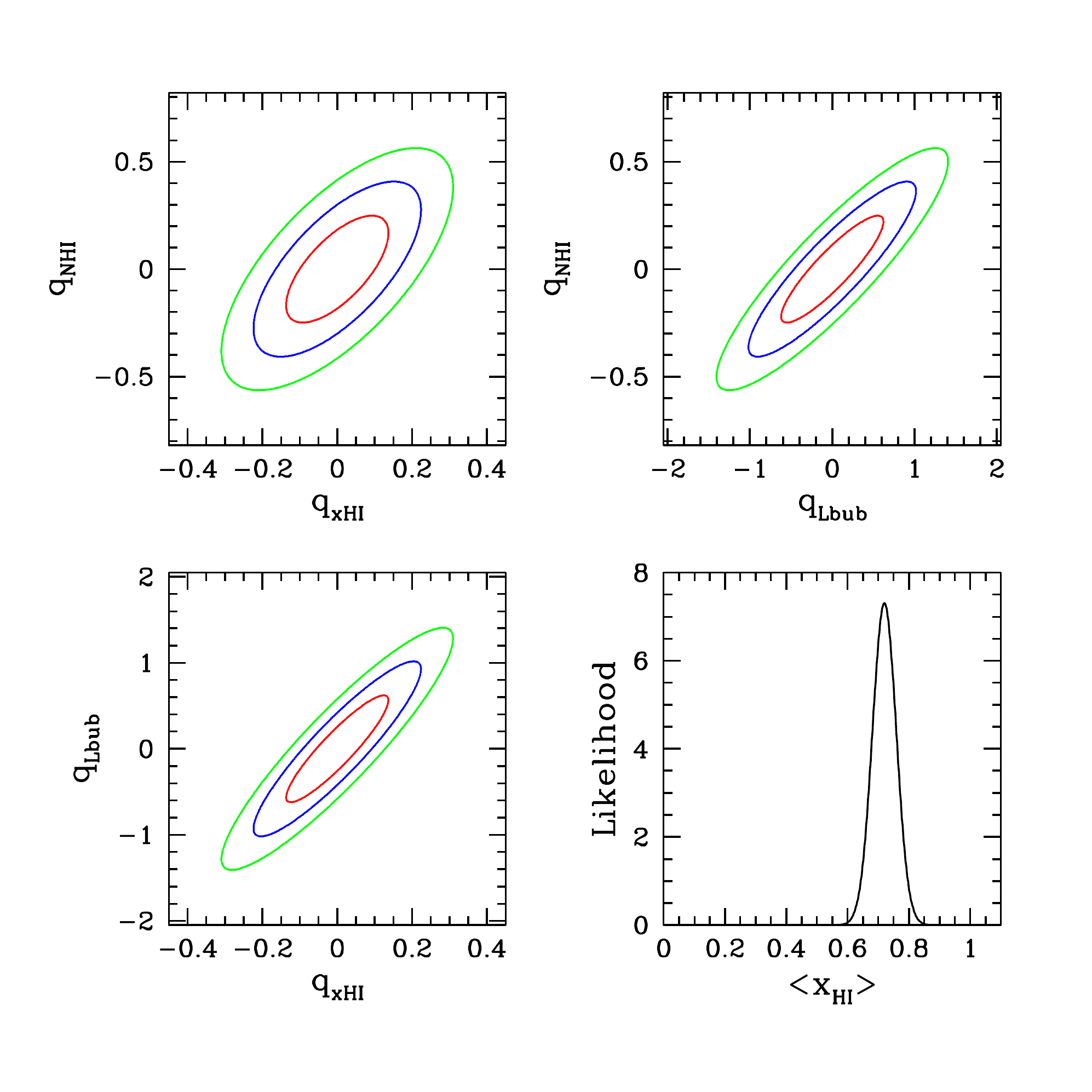}
\caption{Identical to Figure~\ref{fig:param_degen_z6}, but for the $z_{\rm s}=8$ redshift bin.}
\label{fig:param_degen_z8}
\ec
\end{figure}

\begin{figure}[htpb]
\bc
\includegraphics[width=1.0\columnwidth]{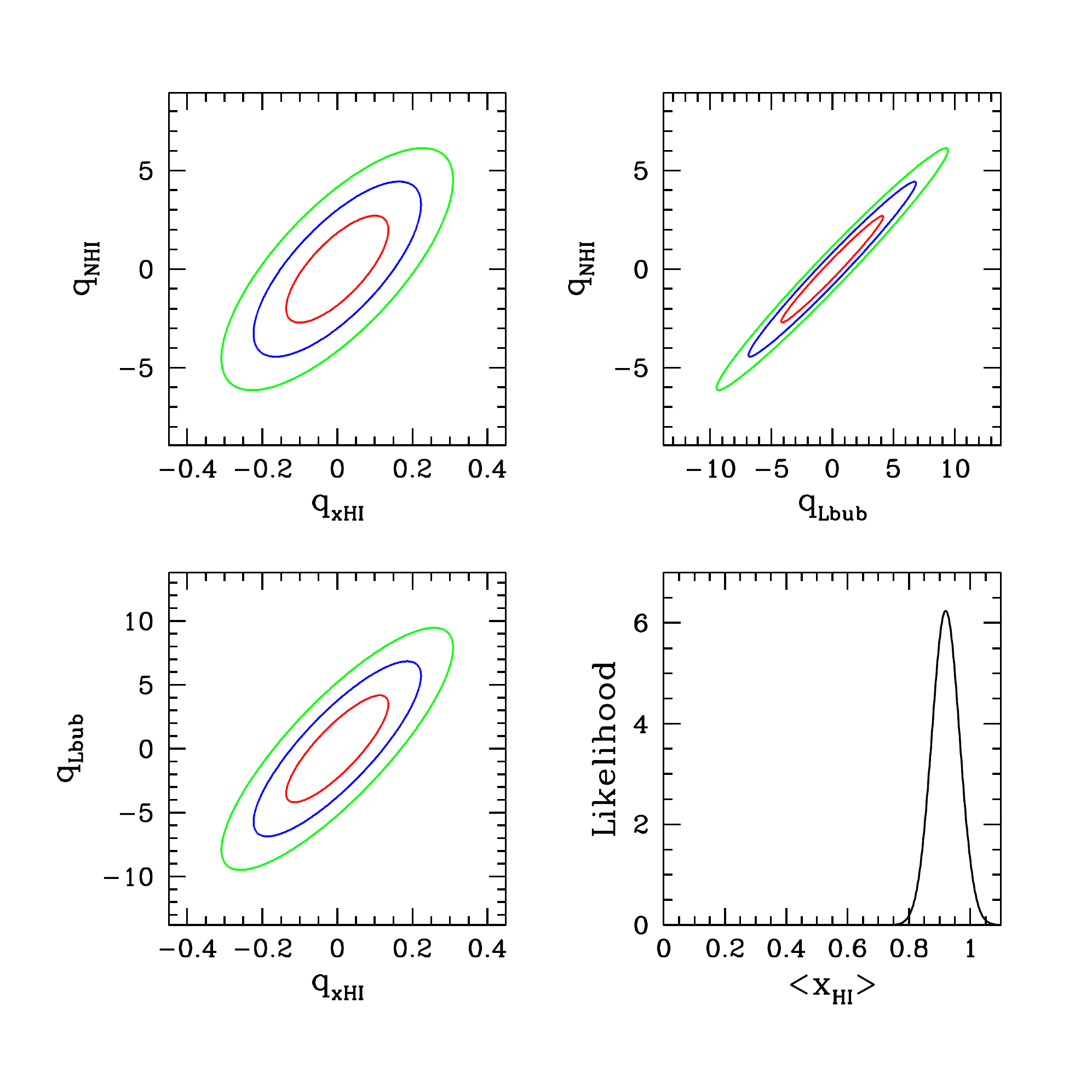}
\caption{Identical to Figures~\ref{fig:param_degen_z6} and \ref{fig:param_degen_z8} but for the $z_{\rm s}=10$ redshift bin.}
\label{fig:param_degen_z10}
\ec
\end{figure}

As an aside, we note that metal absorption lines -- detected slightly redward of the damping wing feature in the same observations -- may help in breaking the degeneracies shown. In particular, low-ionization metal lines should trace the same cold neutral gas in the ISM of the GRB host galaxy that produces the DLA contamination. Specifically, Figure 11 of \cite{MasRibas2016} shows a strong trend between the equivalent width of low-ionization metal absorption lines and DLA column densities, as determined from stacking BOSS quasar spectra. Therefore measurements of the equivalent widths of the low-ionization metal lines may provide independent constraints on the DLA column density parameter towards each GRB afterglow. Although we do not consider this further in what follows, it is an interesting direction for future work. 

Figures~\ref{fig:param_degen_z8} and \ref{fig:param_degen_z10} show analogous plots at $z_{\rm s}=8$ and $z_{\rm s}=10$, respectively. The main change relative to Figure~\ref{fig:param_degen_z6} is that the neutral fraction parameter now has a stronger impact than either of the DLA and bubble size parameters. This reflects the increasing neutral fraction and smaller bubble sizes in our model at these redshifts, which make the IGM damping wing more prominent (see the middle and bottom panel of Figure~\ref{fig:fisher_derivs}). Even though we expect fewer GRB afterglows in these redshift bins than at $z_{\rm s} \sim 6$ (for our fiducial GRB redshift distribution there are $3$ at $z_{\rm s}=8$ and $1$ at $z_{\rm s}=10$, Figure~\ref{fig:grb_histos}), we expect tighter constraints on the neutral fraction. Indeed, the neutral fraction constraints in this simplified case (bottom right hand panels in Figures~\ref{fig:param_degen_z8} and \ref{fig:param_degen_z10}) look promising. In principle, the bubble size parameter itself along with its distribution and redshift evolution, provides valuable information about reionization. However, the error bar forecasts on this parameter are fractionally large (Figures~\ref{fig:param_degen_z6}-\ref{fig:param_degen_z10}), and so we do not consider the reionization information contained in this parameter further in what follows.

\section{Full Ionization History Forecasts}\label{sec:forecasts}

\begin{figure}[htpb]
\bc
\includegraphics[width=1.0\columnwidth]{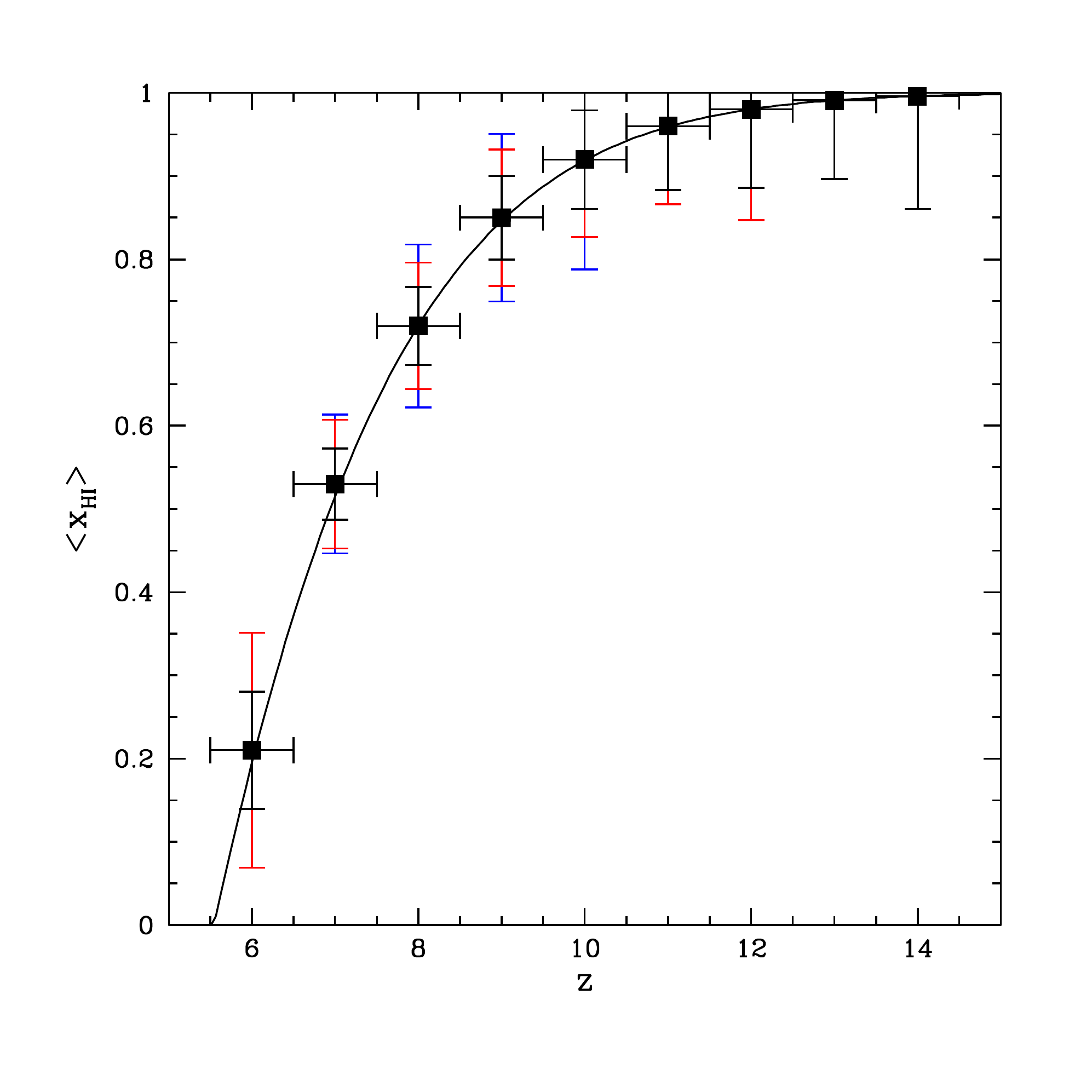}
\caption{Forecasted constraints on the reionization history of the universe. The blue points and 1-$\sigma$ error bars adopt the fiduical GRB redshift distribution of Figure~\ref{fig:grb_histos}, while the red points and error bars show the more optimistic case from the same figure. Finally, the black points and error bars show the constraints for the Theseus-inspired $\sim 80 z \geq 6$ GRBs sample. 
Each of these results incorporate the ionized bubble size distribution and DLA column density PDFs discussed in the text. Each afterglow spectrum is assumed to have an $\rm{SNR}=20$ per $R=3,000$ resolution element at the continuum. Note that in the red/blue models the error bars are identical in the $z_{\rm s}=6$ redshift bin since these cases give an identical GRB count in this bin.}
\label{fig:xhi_of_z}
\ec
\end{figure}

We next compute full ionization history forecasts, assuming the bubble size distribution of Eqs.~\ref{eq:longnorm}-~\ref{eq:sigma_bubble} and the DLA column density PDF of Figure~\ref{fig:pdf_nhi}. The required computations here are described by Eqs.~\ref{eq:fisher_xx_final} and \ref{eq:error_x_final} of Appendix A, and amount to averaging the inverse variance of the neutral fraction over the $L, N_{\rm HI}$ PDFs.

The results of these calculations are shown in Figure~\ref{fig:xhi_of_z}. 
These appear quite encouraging: provided the spectroscopic follow-up observations we considered are feasible, a future GRB mission like the Gamow Explorer should help to determine the redshift evolution of the neutral fraction during cosmic reionization in some detail. Fractionally, the error bar on the neutral fraction is largest towards the end of reionization: in the $z_{\rm s}=6$ bin we forecast a fractional error of 70\% at 1-$\sigma$. In the other redshift bins, the error bars are in the $10-15\%$ range in our fiducial model. The absolute errors range from $\sigma_{\rm xHI} = 0.08-0.14$ with the largest values in the first and last redshift bin.
The error bars are still smaller in the case of the optimistic GRB redshift distribution model, and two additional redshift bins are populated in this case -- out to $z_{\rm s}=12$ -- at which point the fiducial neutral fraction is around $\avg{x_{\rm HI}}=0.98$. 

Finally, the black points and error bars show the more futuristic case in which the Theseus mission obtains high signal-to-noise ratio follow-up spectra for 80 $z \geq 6$ GRB afterglows. In this scenario, the statistical precision of the error bars is further improved, with the error bar forecasts approaching $5\%$ level constraints across many redshift bins in the EoR, degrading mainly towards low redshift (where the damping wing feature is typically somewhat feeble), and at very high redshift where there are few GRBs. Nevertheless, a project like this could potentially determine the neutral fraction from $z_{\rm s} \sim 6-14$ with good statistical precision.

\subsection{Contributions to the Fisher information}\label{sec:fisher_cont}

\begin{figure}[htpb]
\bc
\includegraphics[width=1.0\columnwidth]{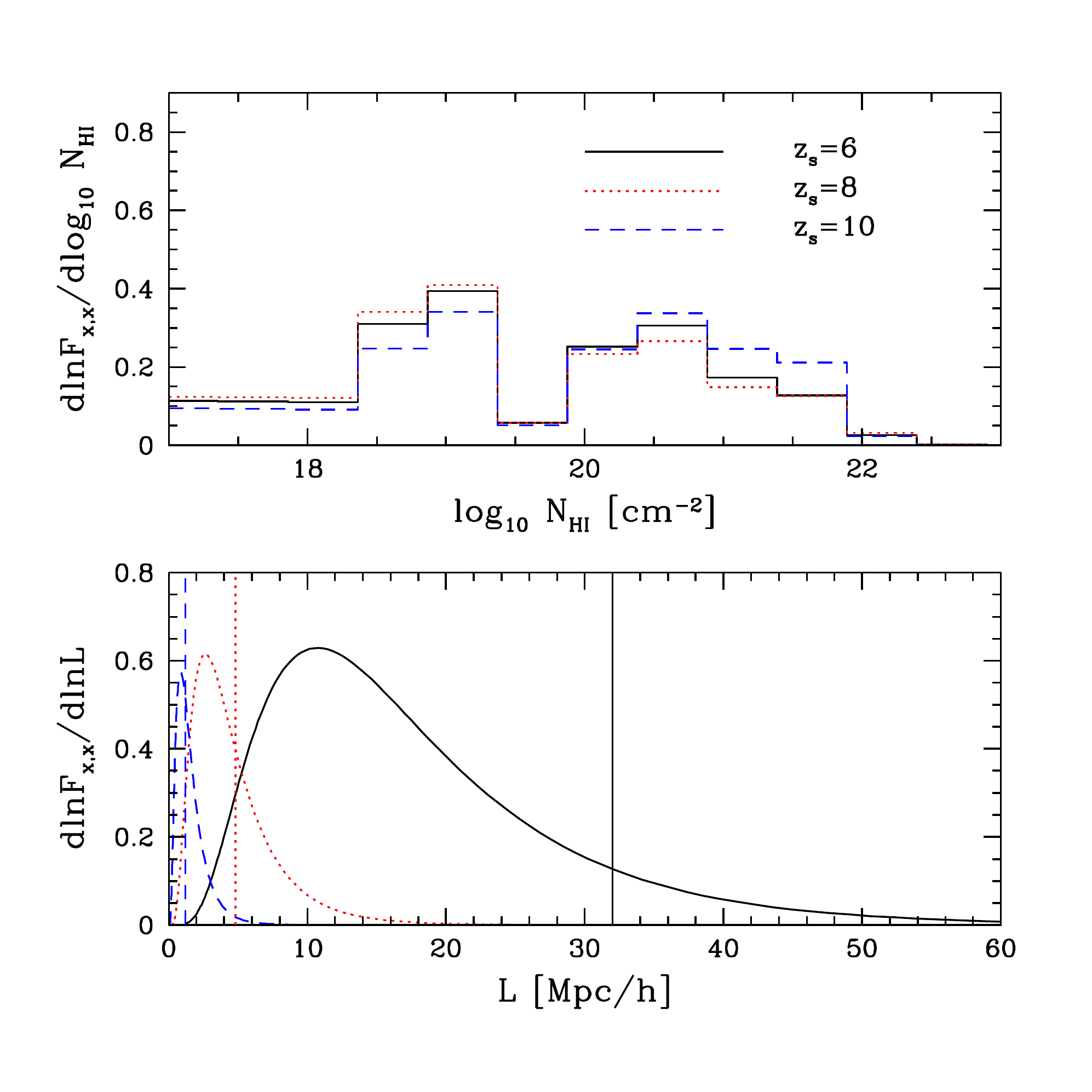}
\caption{Fisher information vs DLA column density and ionized skewer size. {\em Top panel:} The contributions to the neutral fraction Fisher information of Eq.~\ref{eq:fisher_xx_final} as a function of DLA column density at each of $z_{\rm s}=6,8,$ and $10$. The bi-modal nature arises because the $N_{\rm HI}$ PDF is peaked at high column densities (see Figure~\ref{fig:pdf_nhi}), but the sightlines with low column density systems contain more information about the neutral fraction. The histograms are somewhat noisy because they rely on empirical estimates of the $N_{\rm HI}$ PDF. {\em Bottom panel:} The same for the bubble size. The vertical lines indicate the average bubble size at each redshift. The Fisher information here is dominated by sightlines with bubble size smaller than the cosmic mean, with the results turning over as small bubbles bubbles become rare. Here $F_{x,x}$ in the legend is shorthand for $F_{\avg{x_{\rm HI}},\avg{x_{\rm HI}}}.$}
\label{fig:dlnf_xx}
\ec
\end{figure}

In order to unpack these results a little further it helps to examine the integrand of the Fisher information for the neutral fraction in Eq.~\ref{eq:fisher_xx_final}. Specifically, Figure~\ref{fig:dlnf_xx} plots both $d{\rm ln} F_{\avg{x_{\rm HI}},\avg{x_{\rm HI}}}/d {\rm ln} L$
(bottom panel) and $d{\rm ln} F_{\avg{x_{\rm HI}},\avg{x_{\rm HI}}}/d{\rm log}_{10} N_{\rm HI}$ (top panel) as a function of $L$ and ${\rm log}_{10} N_{\rm HI}$, respectively. Put differently, the figure shows how much of the neutral fraction inverse variance is contributed by sightlines with various bubble sizes and column densities. 

The bottom panel illustrates that ionized regions smaller than the mean skewer size dominate the Fisher information at each redshift. The information grows towards small bubble size because the damping wing is stronger for small bubbles, and then turns over as sufficiently small ionized regions become rare. This scale grows as reionization proceeds. 

Likewise, the top panel shows that the Fisher information peaks for rare sightlines with low column density DLAs, although there is still appreciable information in the bins with $20 \lesssim {\rm log}_{10} (N_{\rm HI}/{\rm cm}^{-2}) \lesssim 22$. This reflects a trade-off between the fact that sightlines with lower column densities contain more information about the neutral fraction, but are rarer than the high column density sightlines nearer to the peak of the fiducial $N_{\rm HI}$ distribution (Figure~\ref{fig:pdf_nhi}), which tend to swamp the IGM damping wing signature. The histograms in Figure~\ref{fig:dlnf_xx} are somewhat noisy, because we rely on an empirical model for the $N_{\rm HI}$ PDF. For instance, the dip just below ${\rm log}_{10} (N_{\rm HI}/{\rm cm}^{-2}) = 20$ reflects a corresponding drop in the observed $N_{\rm HI}$ PDF (Figure~\ref{fig:pdf_nhi}). 

Another important implication of this figure is that the constraint forecasts may improve appreciably if the DLA column density PDF evolves strongly with redshift, shifting towards lower column densities, as may be expected if this distribution is directly related to the ionizing photon escape fraction (see \S \ref{sec:dlas}). 
For example, the actual Fisher information in the column density bin around ${\rm log}_{10}(N_{\rm HI}/{\rm cm}^{-2})=18.6$ is a factor of $\sim 7$ larger than the information in the bin centered on  ${\rm log}_{10}(N_{\rm HI}/{\rm cm}^{-2})=20.6$; they only make comparable contributions to $d{\rm ln} F_{\avg{x_{\rm HI}},\avg{x_{\rm HI}}}/d{\rm log}_{10} N_{\rm HI}$ because the lower column density bin here is on the rare tail of the assumed $N_{\rm HI}$ PDF. Note that a factor of $\sim 7$ gain in the Fisher information translates into a smaller error bar on the neutral fraction by a factor of $\sim \sqrt{7.8}=2.8$, and so this could be significant. 

\subsection{Dependence on Spectral SNR}\label{sec:snr_dep}

Here we consider the prospects for obtaining spectral SNRs of $\sim 20$ per $R=3,000$ resolution element at the continuum, as adopted in the calculations of Figure~\ref{fig:xhi_of_z}. We caution that our goal here is only to obtain rough estimates and to explore the dependence on source redshift, flux density, sky background, and observing time. More detailed calculations will be required to cement the observing strategy and improve on our estimates, which are likely accurate only to within a factor of $\sim 2$ in SNR.
We also quantify the impact of varying the SNR around the fiducial value of SNR=20.
To estimate the expected SNR of an observation, we first calculate the number of photons received by a telescope in each spectral resolution element from the source and/or the night sky background:
\beq\label{eq:photon_count}
N_{\rm c, sky} = \frac{f^{\rm c, sky}_{\nu}}{h_p} \epsilon \frac{\pi {\rm D^2_{tel}}}{4}\frac{\Delta \lambda}{\lambda}t_{\rm obs}.
\eeq
Here $N_{\rm c}$ denotes the number of photon counts from the source at the continuum, while $f^{\rm c}_{\nu}$ is the source flux density at the continuum and at an observed wavelength of $\lambda_{\rm obs} = \lambda_\alpha (1 + z_{\rm s})$.\footnote{Throughout this work we quote the SNR at the continuum and near Ly-$\alpha$ at the source redshift. The true SNR in the spectrum will be smaller because of absorption in the damping wing, as properly accounted for in our Fisher matrix calculations.} The quantities $N_{\rm sky}$ and $f^{\rm sky}_\nu$ denote, respectively, the number of photons and the flux density received from the sky background. In this equation, we approximate the telescope aperture as circular, so that the collecting area is $\pi {\rm D^2_{tel}}/4$ where ${\rm D_{tel}}$ is the diameter of the telescope, $\epsilon$ is an efficiency factor to account for photon losses, $\Delta \lambda/\lambda$ is the size of the spectral pixels, $t_{\rm obs}$ is the observing time, and $h_p$ is Planck's constant.We further denote $N_{\rm sky} = \alpha N_{\rm c}$, noting that $\alpha \gg 1$ for ground-based observations. 

In order to estimate the sky background for ground-based observations, we use the VLT X-Shooter exposure time calculator which returns the specific intensity of the sky background in different frequency bands\footnote{\url{https://www.eso.org/observing/etc/bin/gen/form?INS.NAME=X-SHOOTER+INS.MODE=spectro}.  We use the default fixed sky model parameters, a Moon Phase parameter FLI = 0.5, and Airmass = 1.5. We further assume that the sky background is comparable at other good observing sites.}  Specifically, this gives 
$(59, 340, 650) \, \mu {\rm Jy}/{\rm arcsec}^2$ in the (I,Y,J) bands which, respectively, include Ly-$\alpha$ from source redshifts of $z_s=(6,7-8,9-10)$. These values then determine the flux density of the sky background, further assuming one-arcsecond spatial pixels for ground-based observations. It should be possible to improve on this seeing-limited case using adaptive optics, but we do not consider this further in what follows. The sky background is dominated by atmospheric emission, or ``airglow'', and is a strong function of wavelength; this makes the highest redshift GRB measurements challenging from the ground. From space, with the JWST, the sky background is expected to be dominated by Zodiacal light (see below) and such observations will be in the source-dominated regime, i.e.,  $N_{\rm c} \gg N_{\rm sky}$, $\alpha \sim 0$. 

Using Eq.~\ref{eq:photon_count} we can estimate the spectral SNR (squared), accounting for Poisson fluctuations in the source and sky background photon counts as:
\beq\label{eq:snr}
{\rm SNR}^2 = \frac{1}{\alpha + 1} \frac{f^{\rm c}_\nu}{h_p} \epsilon \frac{\pi {\rm D^2_{tel}}}{4}\frac{\Delta \lambda}{\lambda} t_{\rm obs}.
\eeq
We neglect read-out noise and dark current.
Plugging in some characteristic numbers, we find:
\begin{align}\label{eq:snr_numbers}
{\rm SNR}^2 = \, &  20^2 \left[\frac{f^{\rm c}_\nu}{10\, \mu {\rm Jy}}\right] \left[\frac{35}{1+\alpha}\right] \left[\frac{\epsilon}{0.25}\right]
\left[\frac{\Delta \lambda/\lambda}{1/3,000}\right] \nonumber \\
& \times \left[\frac{\rm D_{tel}}{8\, {\rm m}}\right]^2 \left[\frac{t_{\rm obs}}{6.1\, {\rm hrs}}\right].
\end{align}
That is, we find that an 8-meter telescope can achieve an SNR of 20 on a $10\, \mu {\rm Jy}$ GRB afterglow in $t_{\rm obs} = 6.1$ hours of observing time at a spectral resolution of R=3,000. Here we use $\alpha=34$ appropriate for ground-based observations in the Y-band of a source at $z_{\rm s} \sim 8$, where $f^{\rm sky}_\nu = 340 \,\mu {\rm Jy}$.\footnote{We tested the accuracy of Eq.~\ref{eq:snr_numbers} by comparing with the afterglow spectrum in \cite{Totani:2005ng}. This was a $2.6 \mu {\rm Jy}$ burst at $z=6.3$, observed for 4 hours. Adopting a sky background estimate of $\sim 80 \, \mu {\rm Jy}$, in between the I and J band values above, we estimate an SNR of $8$ from Eq.~\ref{eq:snr_numbers} which appears consistent with the published spectrum at the continuum near Ly-$\alpha$.} In other words, an SNR of 20 could be achieved on a $10 \mu {\rm Jy}$ afterglow at $z_{\rm s}=8$ from the ground on an 8-meter telescope in a single night of observations. Although this is a fairly bright burst, note that the \cite{Totani:2005ng} spectrum, for example, was observed $3.4$ days after the burst, and so could have been detected as a $10 \mu {\rm Jy}$ afterglow if it had been followed-up more promptly, a little less than a day after the burst (assuming a typical $\sim t^{-1}$ decay). This can also be seen directly in the light curve of this source presented in the discovery paper by \citet{Haislip:2005mr}.

Moreover, in the future it will be possible to exploit the power of 30-meter class telescopes on the ground and the JWST in space. For example, Eq.~\ref{eq:snr_numbers} shows that using a 30-meter telescope one could observe a burst comparable to the \cite{Totani:2005ng} one, with a flux of $f^{c}_\nu = 3 \, \mu {\rm Jy}$, at an SNR of 20 in 4.7 hours, assuming the $z_{\rm s}=8$ sky background numbers. As mentioned earlier, adaptive optics might help to reduce the sky bacgkround contribution and the required observing time. Using the JWST (with $D_{\rm tel} = 6.5 {\rm m}$), a $f^{c}_\nu = 2 \, \mu {\rm Jy}$ burst observation would reach an SNR of 20 after 1.3 hours of observing time.\footnote{For the case of JWST, we consider NIRSPEC observations with a pixel size of $0.01$ arcsec$^2$ and the highest spectral resolution mode with $R=2,700$. We use the sky background data from \url{https://jwst-docs.stsci.edu/jwst-observatory-functionality/jwst-background-model}. We assume the same efficiency factor, $\epsilon=0.25$, as for the ground-based observations.} Since the JWST observations are source-count limited -- and not by the wavelength-dependent sky background -- the SNR here only depends on the source flux density, with time dilation also counteracting the dimming from increasing luminosity distance towards higher redshifts, as mentioned earlier. JWST will therefore play a crucial role in obtaining high SNR afterglow spectra of the most distant GRBs at $z \gtrsim 9$, where airglow makes ground-based observation especially challenging, although the prospects here with 30 meter telescopes towards relatively bright afterglows are also fairly promising.  

Table~\ref{tab:snr_table} provides a further summary of the exposure time requirements for obtaining an SNR of 20. Here we show examples at 
redshifts $z=6,\, 8$ and 10, for afterglows with specific flux densities of $f_\nu^{\rm c} = 1,\, 5$ and $10\,{\rm \mu Jy}$. We show results for 8 and 30 meter telescopes from the ground, and for space-based observations with the JWST.

While these estimates suggest there are multiple paths forward for obtaining the requisite ${\rm SNR \sim 20}$ follow-up spectra, further work is needed to refine our SNR estimates, and to assess what fraction of the GRBs detectable with the Gamow Explorer, HiZ-GUNDAM, and the Theseus mission will meet these demands. This depends on the flux distribution of observable GRBs and on how rapidly the follow-up spectroscopic observations can be executed.

\begin{figure}[htpb]
\bc
\includegraphics[width=0.96\columnwidth]{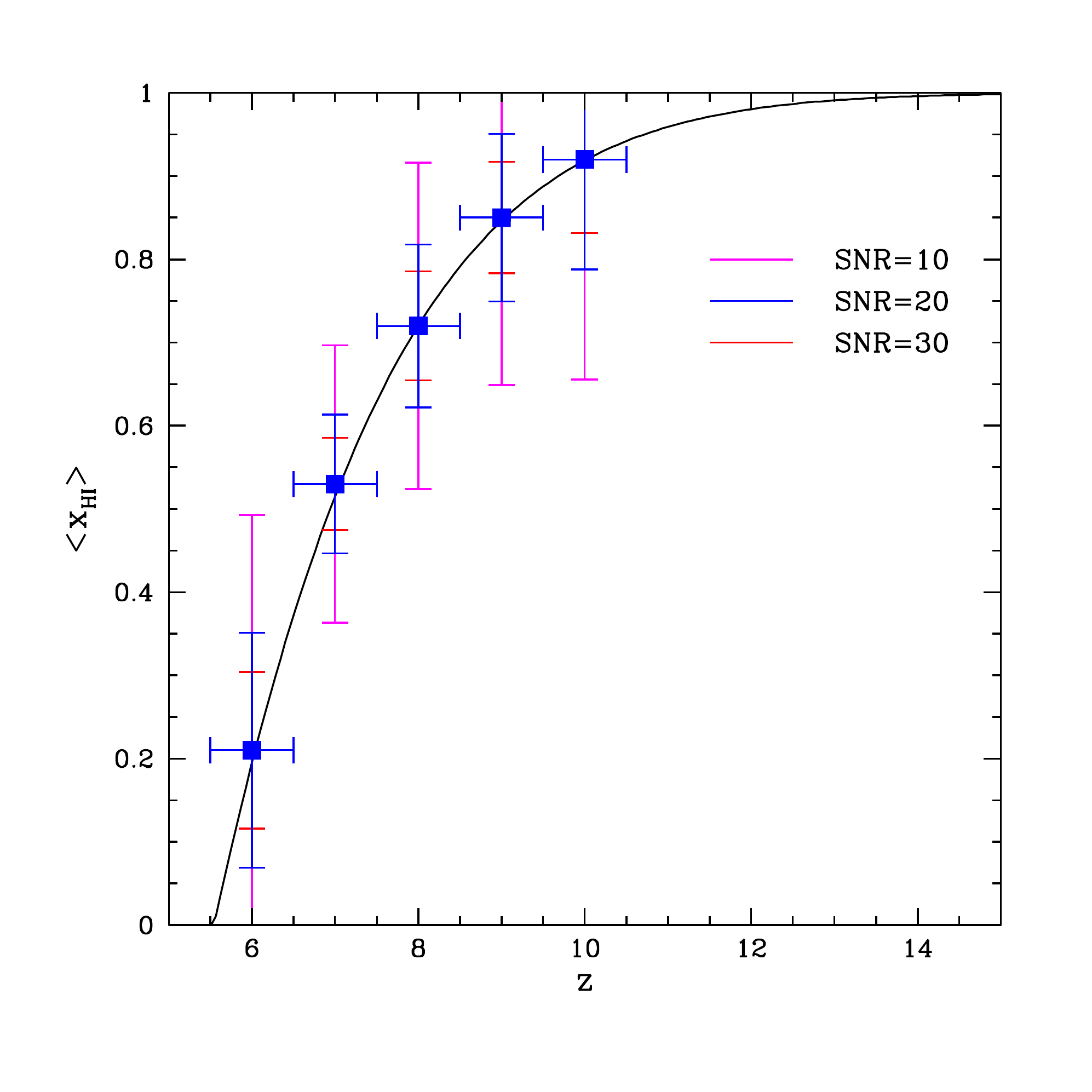}
\caption{Forecasted constraints on the reionization history of the universe vs spectral signal-to-noise-ratio (SNR). Identical to Figure~\ref{fig:xhi_of_z}, except here we vary the SNR of the spectral pixels from SNR=$10$ (magenta error bars) to SNR=$20$ (blue points, identical to the blue points in Figure~\ref{fig:xhi_vs_z}), and SNR=$30$ (red points). In all cases, we assume a spectral resolution of $R=3,000$ and the fiducial GRB redshift distribution.}
 \label{fig:xhi_of_z_snr}
\ec
\end{figure}

We can also illustrate how our Fisher matrix forecasts depend on the quality of the afterglow follow-up observations. This is shown in Figure~\ref{fig:xhi_of_z_snr} and demonstrates the improvements that may be possible if the SNR is boosted to 30. It also shows how much the constraints degrade if the typical follow-up spectrum has SNR=10 per $R=3,000$ resolution element. 

\begin{table*}\center
	\begin{center}
	\caption{Exposure time required to obtain an SNR $=20$ at rest-frame $1216 \Ang$ in the GRB afterglow spectra. These are only intended as rough estimates.}\label{tab:snr_table}
	\begin{threeparttable}
		\begin{tabular}{llllllllll} 
		\hline	
		 $z$           &\multicolumn{3}{c}{$6$}       &\multicolumn{3}{c}{$8$} &\multicolumn{3}{c}{$10$}     \\ 
		 $f_\nu^{\rm c}$       &$1\,{\rm \mu Jy}$   &$5\,{\rm \mu Jy}$ &$10\,{\rm \mu Jy}$           &$1\,{\rm \mu Jy}$   &$5\,{\rm \mu Jy}$ &$10\,{\rm \mu Jy}$        &$1\,{\rm \mu Jy}$   &$5\,{\rm \mu Jy}$ &$10\,{\rm \mu Jy}$      \\ 
		 \hline
		 $f_\nu^{\rm sky}$ (ground)        &\multicolumn{3}{c}{$59\,{\rm \mu Jy\, (I\, band)}$}       &\multicolumn{3}{c}{$340\,{\rm \mu Jy\, (Y\, band)}$}     &\multicolumn{3}{c}{$650\,{\rm \mu Jy\, (J\, band)}$}     \\ 
		8 m	       &$4.4\,{\rm days}$     	&$4.5\,{\rm hours}$			&$1.2\,{\rm hours}$       	&$24.8\,{\rm days}$ &$1.0\,{\rm day}$ &$6.1\,{\rm hours}$ &$47.3\,{\rm days}$   		&$1.9\,{\rm days}$	&$11.2\,{\rm hours}$  \\  
		30 m           &$7.5\,{\rm hours}$     	&$19.2\,{\rm minutes}$			&$5.1\,{\rm minutes}$       	&$1.8\,{\rm days}$ &$1.7\,{\rm hours}$ &$26.0\,{\rm minutes}$          &$3.4\,{\rm days}$   &$3.2\,{\rm hours}$	&$47.7\,{\rm minutes}$   \\ 
				\hline   
		$f_\nu^{\rm sky}$ (space)   &\multicolumn{3}{c}{$0.07\,{\rm \mu Jy}$}       &\multicolumn{3}{c}{$0.07\,{\rm \mu Jy}$}          &\multicolumn{3}{c}{$0.06\,{\rm \mu Jy}$}     \\ 
		6.5 m JWST    &$2.4\,{\rm hours}$ 	&$28.4\,{\rm minutes}$	&$14.2\,{\rm minutes}$       	&$2.4\,{\rm hours}$ 	&$28.4\,{\rm minutes}$	&$14.2\,{\rm minutes}$      &$2.4\,{\rm hours}$ 	&$28.4\,{\rm minutes}$	&$14.2\,{\rm minutes}$   \\  		
		\hline
		\end{tabular}
	\end{threeparttable}
	\end{center}
\end{table*}

\section{Constructing a census of the ionizing sources}\label{sec:census}

In this section we briefly illustrate how the combination of GRB afterglow constraints on both the escape fraction and the ionization history can help assess whether we have an accurate census of the ionizing source populations. 
The redshift distribution of long-duration GRBs also offers a handle on the evolution of the SFRD -- provided evolution in the GRB to star-formation rate is understood and/or empirically well calibrated. Essentially, the source term in Eq.~\ref{eq:xofz} is determined (at each redshift) by the SFRD, the escape fraction, and the ionizing spectrum of the sources. The recombination term in Eq.~\ref{eq:xofz} is generally less important, since the recombination time is fairly long near the cosmic mean density and current simulation work suggests a fairly small clumping factor, $C \sim 2-3$ (e.g., \citealt{McQuinn2011}), during most of the EoR.
Hence if the ionizing spectrum may be constrained with e.g., JWST observations, GRB afterglow data may allow one to predict the ionization history from the observed SFRD and escape fraction; this may be compared with the damping wing measurements of the reionization history.\footnote{One caveat here is that Eq.~\ref{eq:xofz} assumes a simplified form for the source term, where for instance the ionizing efficiency factor is independent of host halo mass and redsfhit. A more complex model will be required to interpret the data, but we focus on this simple model here for the purpose of illustration.} The SFRD determinations may of course be compared with that from JWST observations, which probe the bright end of the galaxy populations. Line-intensity mapping offers another potential handle on the cumulative impact of faint source populations \citep{Sun:2020mco,Kovetz:2017agg}. Taken together, these observations should provide a valuable test of whether we have an accurate census of the sources that reionized the universe. 

The escape fraction determinations clearly play a key role here, because this quantity is highly uncertain both theoretically and empirically, and there is relatively little hope (other than through GRB afterglows) for direct constraints during the EoR, especially in the small mass galaxies that may dominate the reionization process. A fairly large sample of afterglows with follow-up spectroscopy is, however, required to determine the escape fraction. The data required motivates follow-up spectroscopy of afterglows at slightly lower redshift, e.g., near $z \sim 5$: the higher SFRD rate and slightly less stringent requirements for detecting these more nearby sources will yield larger samples than the 20 $z_{\rm s} \geq 6$ afterglows expected for the Gamow Explorer. 

This post-reionization sample avoids any damping wing contribution from the IGM; the escape fraction is then determined by fitting for the neutral hydrogen column density associated with each GRB host galaxy, and averaging over the sample of sightlines (e.g., footnote 4, \citealt{Chen:2007wi,Tanvir:2018pbq}). The likely scenario here is that a fraction $1-f_{\rm esc}$ of the sightlines have DLAs and hence effectively vanishing ionizing photon leakage, while $f_{\rm esc}$ of the afterglows show negligible neutral hydrogen absorption, i.e.,  no DLAs or even Lyman-limit systems. In this case, the requirements for follow-up spectroscopy are likely less stringent than in the case of measuring the full shape of the IGM damping wing. In a first pass, one mainly wants to determine whether or not a given sightline has a DLA or not. For estimating the escape fraction, it is not necessary to determine the precise DLA column density since sightlines with DLAs give negligible escape fractions in any case. In the case of sightlines that do not show DLAs, further follow-up may be required to determine whether there is a Lyman-limit system and the associated column density.

\subsection{Prospects for Measuring the Escape Fraction}

For example, suppose that the escape fraction is $f_{\rm esc}=0.1$ and that roughly 100\% of ionizing photons escape along a fraction $f_{\rm esc}$ of sightlines while no ionizing photons escape from the remaining $1-f_{\rm esc}$ of afterglows. In this case, on average one expects $N_{\rm esc} = f_{\rm esc} N_{\rm GRB}$ sightlines to show escaping ionizing photons in a sample of $N_{\rm GRB}$ afterglows. Assuming Poisson fluctuations in $N_{\rm esc}$ around this average, the fractional error on the escape fraction from the sample is $\sigma_{f_{\rm esc}}/f_{\rm esc} \sim 1/\sqrt{N_{\rm esc}} \sim 1/\sqrt{f_{\rm esc} N_{\rm GRB}}$. That is, in a sample of $N_{\rm GRB}=100$ GRB afterglows, we expect $N_{\rm esc}=10$ sightlines to show escaping ionizing photons, and $\sim 30\%$ errors on the escape fraction estimate. Here we assume that the errors are dominated by Poisson fluctuations in the number of sightlines with escaping ionizing photons, and ignore additional errors that may arise from imperfect column density determinations, for example. 
Note that the escape fraction assumed here implies strong redshift evolution from the \cite{Tanvir:2018pbq} sample, but this may be required to reionize the universe with star-forming galaxies (e.g., \citealt{Bouwens2016}), as discussed in \S \ref{sec:dlas}. It would be interesting to split future samples by redshift and stellar mass (obtained from follow-up host galaxy observations) to explore any correlations or trends in these quantities.

\begin{figure}[htpb]
\bc
\includegraphics[width=1.0\columnwidth]{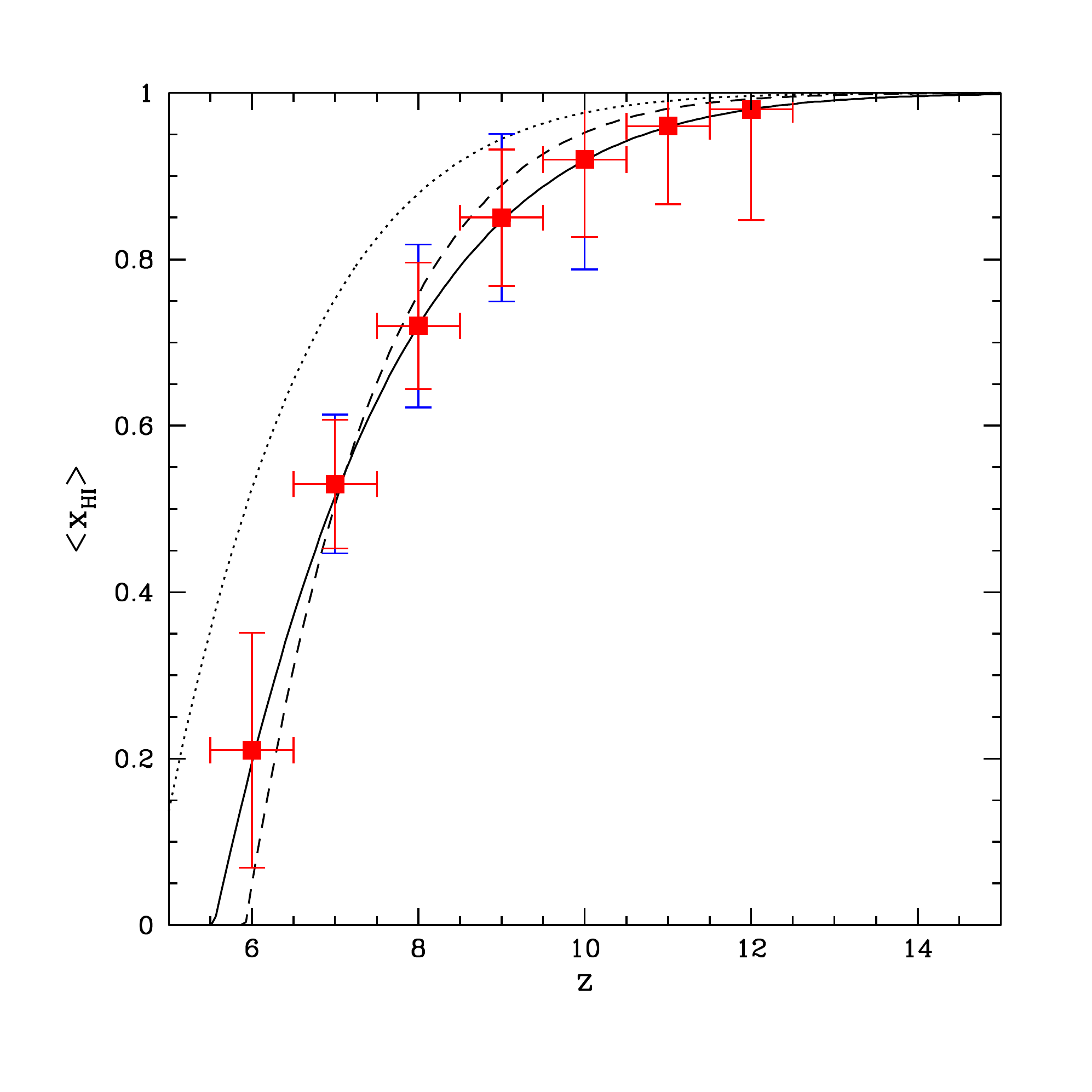}
\caption{Illustration of how joint constraints on the ionization history and escape fraction can improve our understanding of ionizing source
populations. The red and blue points with error bars are the forecasted constraints on the reionization history from Figure~\ref{fig:xhi_of_z}.
The black solid line shows the fiducial model (with a minimum host halo mass of $M_{\rm min} = 10^9 M_\odot$). The black dotted line shows the impact of increasing the minimum host halo mass in the model to $M_{\rm min} = 5 \times 10^9 M_\odot$, while fixing the escape fraction. The black dashed line also adopts $M_{\rm min} = 5 \times 10^9 M_\odot$, but increases the escape fraction by a factor of two, which largely compensates for the delay in reionization with this $M_{\rm min}$. A measurement of the escape fraction can help rule out this case, which would otherwise be allowed by the ionization history measurements and determinations of the bright end of the UV luminosity function.}
\label{fig:xhi_consistency_test}
\ec
\end{figure}

Even a modest $\sim 30\%$ level constraint on the escape fraction would be interesting and helpful. In order to illustrate this, Figure~\ref{fig:xhi_consistency_test} considers an alternate model for the ionizing sources, in which our fiducial minimum galaxy hosting halo mass is boosted from $M_{\rm min} = 10^9 M_\odot$ to $M_{\rm min} = 5 \times 10^9 M_\odot$. This is intended to loosely mimic scenarios in which there are {\em no additional faint galaxy populations} below the threshold of UV luminosity function measurements: that is, it is a case where bright galaxy populations produce most of the ionizing photons, termed ``reionization by oligarchs'' by \cite{Naidu:2019gvi}. Although different than our fiducial model, this type of scenario is also in agreement with current data, as discussed in that work. Adopting the same value of the escape fraction,  or more precisely the same ionizing efficiency factor in Eq.~\ref{eq:xofz}, gives the dotted black curve in Figure~\ref{fig:xhi_consistency_test}. This delays the reionization history of the universe relative to our fiducial model and would be ruled-out by the damping wing measurements given the forecasts of the previous section. In fact, this case is already moderately disfavored by current observations (e.g., Figure~\ref{fig:xhi_vs_z}), but suffices for illustration. 

However, one can compensate for this delay by increasing the escape fraction. In particular, the dashed line in Figure~\ref{fig:xhi_consistency_test} boosts the escape fraction by a factor of two and this largely counteracts the effects from increasing the minimum mass.\footnote{Formally, note that the dashed line case would also be disfavored at moderate statistical significance by the ionization history constraints. We sharpen this in the Fisher analysis below, \S \ref{sec:fisher_source}.}
Put differently, there are degeneracies between the minimum host halo mass and the escape fraction (see also, for example, Figure~9 of \citealt{SF_2016}). This can be broken by adding the GRB determinations of the escape fraction, which -- in our fiducial model -- would argue against the factor of two boost in escape fraction. Without the escape fraction measurements, one could be spuriously led into thinking that the brighter sources reionize the universe. Hence the combination of ionization history constraints and escape fraction measurements can play an important role in determining whether we have accurately identified the source populations responsible for reionization. Furthermore, the comparison between the redshift distribution of the GRBs and the SFRD determined from bright populations with e.g., the JWST might provide independent evidence for important contributions from faint source populations, since long-duration GRBs may be produced in such galaxies, while these may remain below the JWST flux limits.

\subsection{Source Parameter Constraints}\label{sec:fisher_source}

\begin{figure}[htpb]
\bc
\includegraphics[width=1.0\columnwidth]{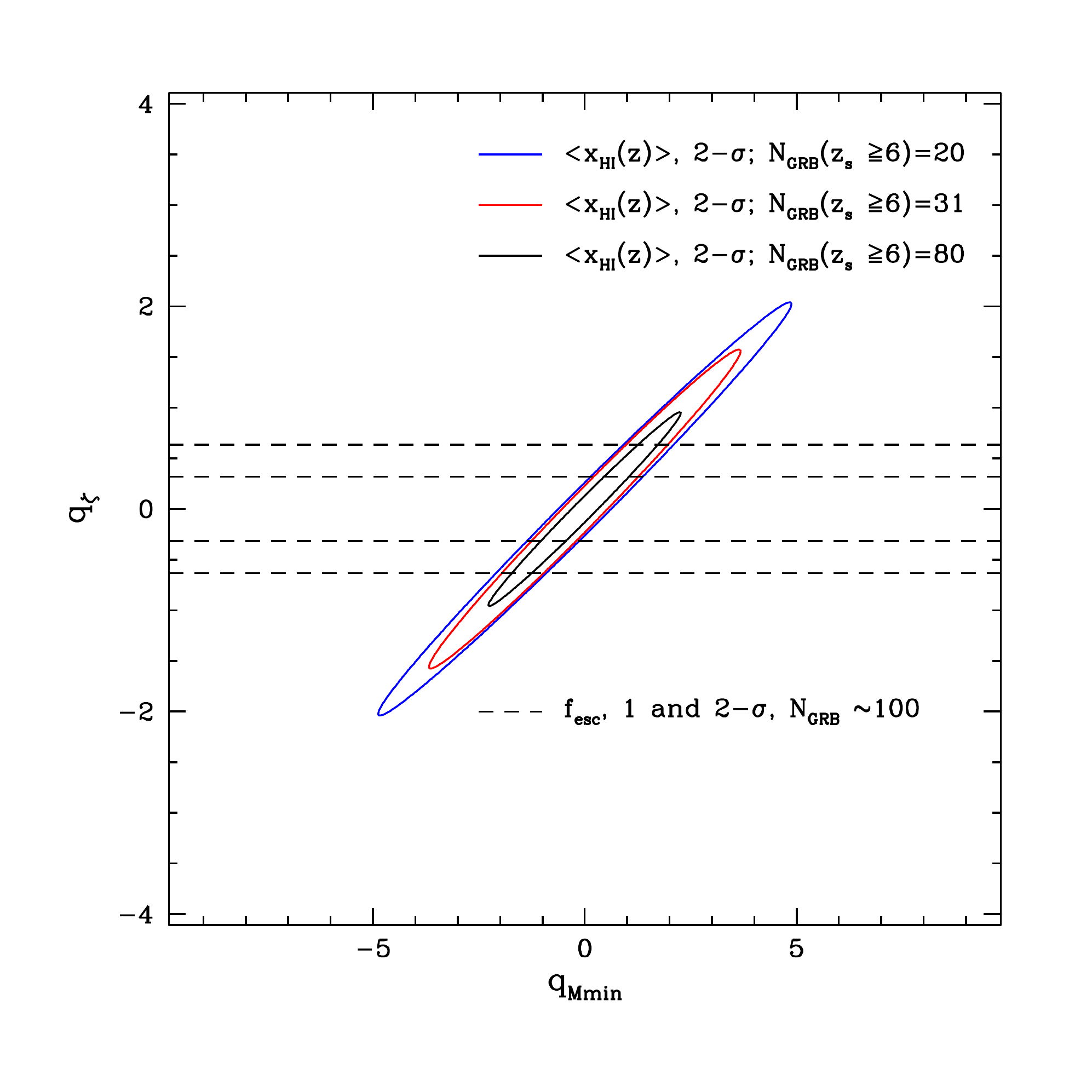}
\caption{Constraints on the ionizing source properties from the reionization history and escape fraction measurements. The blue, red, and black contours show $2-\sigma$ constraints on the ionizing efficiency parameter, $\zeta$, and the minimum host halo mass of the ionizing sources, $M_{\rm min}$, from the ionization history constraints for the fiducial, optimistic, and Theseus-inspired GRB samples, respectively. The black dashed lines show 1 and $2-\sigma$ constraints on the escape fraction from a sample of 100 afterglows, assuming that the uncertainty in the ionizing efficiency parameter is dominated by that in the escape fraction (see text). The combination of these measurements is helpful for determining the $M_{\rm min}$ parameter.}
\label{fig:contour_source}
\ec
\end{figure}

We can sharpen this argument by further forecasting the error bars on the parameters of the ionizing sources, which may be obtained from both the GRB afterglow constraints on the ionization history and the escape fraction determinations. To do this, we perform a Fisher matrix forecast for the ionizing source parameters in Eq.~\ref{eq:xofz}, given the error bars previously forecast for $\avg{x_{\rm HI}(z)}$. Here, for simplicity, we consider a two-parameter model, characterized by $q_\zeta = (\zeta-\zeta_{\rm fid})/\zeta_{\rm fid}$ and $q_{\rm Mmin} = (M_{\rm min} - M_{\rm min,fid})/M_{\rm min,fid}$. That is, we fix the clumping factor at $C=3$ and ignore any redshift or mass dependence in these parameters. Our fiducial parameter values are $\zeta_{\rm fid}=20$ and $M_{\rm min}=10^9 M_\odot$. The Fisher matrix for these parameters may be computed as
\beq\label{eq:fisher_source}
F^\prime_{ij}=\sum_k \frac{\partial \avg{x_{\rm HI}(z_k)}}{\partial q_i}\frac{\partial \avg{x_{\rm HI}(z_k)}}{\partial q_j}\frac{1}{{\rm var}\left[\avg{x_{\rm HI}(z_k)}\right]},
\eeq
where the prime notation distinguishes this Fisher matrix from the one in Eq.~\ref{eq:fisher_conditional}. The sum runs over the redshift bins. The parameter derivatives are computed from the ionization equation (Eq.~\ref{eq:xofz}), and the variance of the neutral fraction estimates in each bin follow from the calculations in \S \ref{sec:forecasts}. Finally, we can add the independent information provided by an escape fraction estimate. Note that in principle the ionizing efficiency parameter depends on the overall star-formation efficiency and the ionizing spectrum, as well as the escape fraction. For simplicity, we suppose that in the JWST-era the escape fraction is the main unknown, so that a $\sim 30\%$ error in the escape fraction translates into a $\sim 30\%$ error in the ionizing efficiency. In this case, assuming a fiducial escape fraction of $10\%$ and a sample of 100 GRBs determines $\zeta$ with a fractional accuracy of $\sim 1/\sqrt{10}$ as discussed earlier.

The resulting constraint ellipses are shown in Figure~\ref{fig:contour_source} for each of the GRB redshift distributions in Figure~\ref{fig:grb_histos}, which result in the corresponding neutral fraction error bars in Figure~\ref{fig:xhi_of_z}.\footnote{As in our previous results, we do not adopt positivity priors on the source parameters and so $q_i \leq -1$ in some cases.} As anticipated previously, the ionization history measurements alone leave a strong positive degeneracy between $q_\zeta$ and $q_{\rm Mmin}$. However, the $f_{\rm esc}$ measurement can help break this degeneracy. Quantitatively, in our baseline $N_{\rm GRB}(z \geq 6) = 20$ model, we find that the fractional $M_{\rm min}$ error is 74\% for a joint constraint but only 200\% without the information from $f_{\rm esc}$. For the more optimistic GRB redshift distribution with $N_{\rm GRB}(z \geq 6)=31$, the corresponding numbers are 69\% and 150\%. In the Theseus-inspired scenario, with $N_{\rm GRB}(z \geq 6)=80$, these numbers improve to 59\% and 92\%, respectively.\footnote{In the Theseus case, one might imagine that this more futuristic mission could obtain a larger set of afterglows for the escape fraction measurement, which would tighten these limits.} A good deal of the constraining power is driven by the bursts at high redshift: for instance, if we only include the bursts at $z \geq 9$ in the Theseus case, the error bars on the minimum mass only increase by $8\%$. The high redshift bursts play a valuable role here because the halo mass function becomes exponentially sensitive to $M_{\rm min}$, as the galaxy hosting halos form only in the rare high-$\sigma$ peaks of the mass distribution in these highest redshift bins. 
The numbers quoted are $1-\sigma$ constraints. Hence the escape fraction limits do help noticeably, and they may be still more valuable if $f_{\rm esc}$ is larger than our fiducial value of $f_{\rm esc}=0.1$, which would improve the escape fraction measurement error. 

Further, we have adopted a very simple model for the ionizing sources in this illustration. In more realistic cases the parameters may depend on halo mass, redshift, or other galaxy host properties. It will be important to combine ionization history constraints with a full suite of measurements to help pin down the properties of the ionizing sources. Determinations of the escape fraction, and any measurements or limits on the redshift dependence of this quantity and/or correlations with host properties should be extremely valuable here. Although our simple illustration shows improvements on the $M_{\rm min}$ constraint from an $f_{\rm esc}$ determination, this may in fact understate the value of such measurements.

\section{Conclusions}\label{sec:conclusions}

We have forecast the constraints on the reionization history for upcoming GRB afterglow samples during the EoR. Assuming 20 $z \geq 6$ GRB afterglow spectra with a typical SNR of 20 per $R=3,000$ spectral resolution element at the continuum, we find that the volume averaged neutral fraction can be determined to 10-15\% accuracy from $z \sim 7-10$, and with $\sim 70\%$ errors near $z=6$. Although there appear to be reasonable prospects for obtaining the requisite follow-up observations (see \S \ref{sec:snr_dep}), more work is required to assess the number of GRBs that will meet these requirements in different redshift bins. 

The ionization history forecasts represent a significant improvement on current constraints (e.g., Figure~\ref{fig:xhi_vs_z}), and will provide a useful cross-check on future damping wing measurements towards high redshift quasar samples. Perhaps more importantly, GRBs can help extend the quasar measurements past $z \gtrsim 8-9$ to probe earlier phases of the reionization history before bright quasars formed. Measurements of the redshifted 21 cm power spectrum and its evolution may offer more precise measurements of the reionization history \citep{DeBoer:2016tnn} than our GRB forecasts, but these rely on understanding the mapping between the observable, spatial fluctuations in the 21 cm signal and the average ionization history, and face a range of systematic challenges \citep{Liu:2019awk}. While the GRB constraints also rely on understanding the mapping between an observable (the damping wing) and the average ionization history, they provide an independent constraint from the 21 cm efforts, and will be valuable. In the future, it will be interesting to consider joint fits with 21 cm, future cosmic variance limited CMB EE polarization measurements (which provide mostly an integral constraint on the ionization history, e.g., \citealt{Heinrich:2016ojb}), the patchy kinetic Sunyaev-Zel'dovich effect \citep{Ferraro:2018izc}, the quasar constraints, Lyman-alpha emitter samples (e.g., \citealt{Mason:2019ixe}), and other probes. 
In the slightly longer term, a sample of as many $\sim 80$ $z \geq 6$ GRB afterglows from the Theseus mission \citep{Amati:2017npy}, with follow-up spectroscopy, could return constraints as tight as $\sim 5\%$ across much of the EoR and place some limits out to $z \sim 14$. In principle, these measurements could be used to infer the electron scattering optical depth to CMB photons, and thereby sharpen cosmological parameter constraints (e.g. \citealt{Liu:2015txa}). 

We emphasized that an appealing aspect of GRB afterglow probes of reionization is that these constraints may also be combined with escape fraction determinations from the GRB column density distribution and with internal or external measurements of the SFRD. In order, however, to build up a sufficient afterglow sample to determine the escape fraction we recommend a follow-up campaign targeting afterglows at slightly lower redshifts, $z \sim 5$ and $50-100$ afterglows. Taken together, the ionization history, escape fraction, and SFRD constraints will provide a powerful test of our census of the sources that reionized the universe.

It will also be interesting to perform follow-up observations to compare the large-scale environment around GRBs with strong and weak damping wings. Surveys of the galaxy distribution and/or line-intensity maps in various emission lines can potentially identify some of the surrounding galaxies responsible for ionizing the intergalactic gas near GRBs that show weak damping wing absorption. It may also be possible to construct maps of the surrounding neutral hydrogen using 21 cm surveys \citep{Beardsley:2014bea}, which would presumably be brighter in the proximity of GRBs showing strong damping wing absorption. The combination of these observations should help in understanding the interplay between the ionizing sources and the neutral gas in the IGM during reionization. 

The biggest caveat with our current work is that we assume a simplified three parameter model for the damping wing signature towards each GRB afterglow. Although patchy reionization is partly incorporated into this model, we assume a uniform neutral fraction outside of the ionized bubble around the GRB host. This description does not fully account for the -- generally complicated -- distribution of neutral and ionized patches towards each afterglow \citep{McQuinn:2007gm,Mesinger:2007kd}. While we expect our simplified description to be mostly adequate for forecasting the constraints that may be obtained with future missions, it is insufficient for interpreting upcoming measurements. The model should be least accurate in the late stages of reionization when the ionized structures become large compared to the effective damping wing length scale. Note that while neglecting scatter in the exterior neutral fraction may cause us to underestimate the error forecasts on this quantity, it also implies that the damping wing feature will be stronger and more detectable towards some lines of sight than in our model; the precise impact of this approximation is unclear and likely sensitive to the reionization model considered. 

We encourage further simulation studies to improve on this treatment. One approach may be to develop a generalized version of the three parameter model informed by mock damping wing spectra extracted from reionization simulations. In this case, a description might include the extent of the first neutral region, the size of the second ionized patch, and so forth. In this case, the average neutral fraction may no longer even be a parameter in the model. Instead, one considers only the length of each neutral segment and their distance to the source. These could be fit for independently of any particular model, while the simulations could be used in a second more model-dependent step to calibrate the relationship of these parameters to the volume averaged neutral fraction. It may also be possible to calculate the average transmission profile around GRB hosts of a given halo mass using a halo-model \citep{Cooray:2002dia} type calculation (see Appendix C of \citealt{Malloy:2014tba} for work in this direction).

The simulations will also enable mock end-to-end tests of parameter extraction pipelines. Such efforts can test the Fisher matrix approximation adopted in this work, which assumes a quadratic expansion in the logarithm of the likelihood around a fiducial input model. A suite of mock spectra may further help in understanding the sample-to-sample variations in the expected error bars; our Fisher forecasts assume typical samples and do not consider the error on the error bar.

Another important issue concerns the redshift evolution of the DLA column density distribution. Our fiducial model adopts the current empirical distribution of DLA host column densities \citep{Tanvir:2018pbq}, but this may be a pessimistic assumption. It might be interesting to use numerical simulations of galaxy formation, capable of modeling the escape fraction of ionizing photons (e.g., \citealt{Ma:2020vlo}), to try and {\em predict} the redshift evolution in the DLA column density distribution towards GRB hosts. Interesting in its own right, this could also be used in our forecasts in place of the empirical distribution, which mostly consists of lower redshift GRBs. 

In summary, the prospects appear promising for future GRB missions to extract valuable constraints on the reionization history. There is a good deal of work to be done to better understand the expected damping wing signatures and to best prepare for the upcoming missions, but we expect these to play an important role in our understanding of cosmic reionization.

\section*{Acknowledgements}

We thank Matt McQuinn for providing the simulation outputs used in computing the ionized skewer size distributions in Figure~\ref{fig:bubble_size}. We thank the Gamow Explorer team for inspiring this work and for useful discussions, especially with Nial Tanvir, Nicholas White, Dieter Hartmann, Joseph Hennawi, Amy Lien, Ruben Salvaterra and Giancarlo Ghirlanda. 
AL acknowledges support, in part, through NASA ATP grant 80NSSC20K0497. Part of this work was done at Jet Propulsion Laboratory, California Institute of Technology, under a contract with the National Aeronautics and Space Administration.

\bibliography{references}

\section*{Appendix A}\label{sec:append_a}

Here we describe how to account for the distribution in ionized skewer size and the DLA column density PDF in our Fisher matrix calculations.
Let us first consider the conditional likelihood of the afterglow transmission data for a sightline with a given bubble size and column density, denoted by $\mathcal{L}({\bf \hat{f}, q}|L, N_{\rm HI})$, with 
\beq\label{eq:likelihood_conditional}
-{\rm ln} \mathcal{L}({\bf \hat{f},q}|L, N_{\rm HI}) = \sum_k  \left[\hat{f}(k) - f(k, {\bf q})\right]^2 \frac{1}{{\rm var}\left[f(k)\right]},
\eeq
where we have assumed a Gaussian likelihood. Here $\hat{f}$ denotes a measured transmission, while $f({\bf q})$ describes a transmission model, and the sum is over independent spectral pixels.
The conditional Fisher matrix, denoted $\tilde{F}_{ij}(L,N_{\mathrm{HI}})$, then follows from the definition of the Fisher matrix as
\beq\label{eq:fisher_cond_def}
\tilde{F}_{ij}(L,N_{\rm{HI}}) = - \left\langle \frac{\partial^2 {\rm ln} \mathcal{L}({\bf \hat{f},q}|L, N_{\rm HI})}{\partial q_i \partial q_j} \right \rangle,
\eeq
and this results in Eq.~\ref{eq:fisher_conditional} in the main body of the text.

It is further useful to define the likelihood of a neutral fraction estimate, denoted $\hat{x}$ here for brevity of notation, for a given realization of bubble size and DLA column density, yet marginalized over the estimates of bubble size and column density. We denote this quantity as $\mathcal{L}(\hat{x}|L, N_{\rm HI})$. We can compute the width of this distribution, around some fiducial volume-averaged neutral fraction, by determining the inverse of the conditional Fisher matrix. Specifically,
\beq\label{eq:varx_cond}
\rm{var}\left[\avg{x_{\rm HI}}|L,N_{\rm HI}\right] = \left[\left\langle \frac{\partial^2 {\rm ln} \mathcal{L}(\hat{x}|L,N_{\rm HI})}{\partial \hat{x}^2}\right\rangle \Bigr|_{\hat{x}=\avg{x_{\rm HI}}}\right]^{-1} = \avg{x_{\rm HI}}^2 \left[\tilde{F}^{-1}\right]_{q_{\rm xHI}, q_{\rm xHI}},
\eeq
where the factor of $\avg{x_{\rm HI}}^2$ in the second equality is needed to move from the variance of $q_{\rm xHI}$ (see \S \ref{sec:fisher}) to that of $\avg{x_{\rm HI}}$ itself.

Next, we can finally incorporate the bubble size and column density distributions. The likelihood of $\hat{x}$, marginalized over the bubble size and column density PDFs is:
\beq\label{eq:lx_marg}
\mathcal{L}(\hat{x}) = \int dL dN_{\rm HI} \frac{dP_L}{dL} \frac{dP_{\rm NHI}}{dN_{\rm HI}} \mathcal{L}(\hat{x}|L, N_{\rm HI})
\eeq
The {\em non-conditional} Fisher matrix of this one-parameter distribution is given by:
\beq\label{eq:fxx}
F_{\avg{x_{\rm HI}}, \avg{x_{\rm HI}}} = - \left\langle \frac{\partial^2 {\rm ln} \mathcal{L}(\hat{x})}{\partial \hat{x}^2}\right\rangle \Bigr|_{\hat{x}=\avg{x_{\rm HI}}},
\eeq
which is just the inverse variance of the $\hat{x}$ distribution. We can evaluate this expression and relate the results to the
conditional variance in Eq.~\ref{eq:varx_cond}. First note that
\beq\label{eq:lnderiv}
\frac{\partial {\rm ln} \mathcal{L}(\hat{x})}{\partial \hat{x}} = \int dL dN_{\rm HI} \frac{dP_L}{dL} \frac{dP_{\rm NHI}}{dN_{\rm HI}}
\frac{\mathcal{L}(\hat{x}|L,N_{\rm HI})}{\mathcal{L}(\hat{x})} \frac{\partial {\rm ln} \mathcal{L}(\hat{x}|L, N_{\rm HI})}{\partial \hat{x}}
\eeq
We take the second derivative of Eq.~\ref{eq:lnderiv}, using that $\partial \mathcal{L}(\hat{x}|L, N_{\rm HI})/\partial \hat{x} |_{\hat{x}=\avg{x_{\rm HI}}} = \partial \mathcal{L}(\hat{x})/\partial \hat{x} |_{\hat{x}=\avg{x_{\rm HI}}} = 0$. This gives:
\beq\label{eq:lnderiv_two_a}
-\left\langle \frac{\partial^2 {\rm ln} \mathcal{L}(\hat{x})}{\partial \hat{x}^2} \right\rangle \Bigr|_{\hat{x}=\avg{x_{\rm HI}}} = -\int d\hat{x} dL dN_{\rm HI}  \mathcal{L}(\hat{x}) \frac{dP_L}{dL} \frac{dP_{\rm NHI}}{dN_{\rm HI}}
\frac{\mathcal{L}(\hat{x}|L,N_{\rm HI})}{\mathcal{L}(\hat{x})} \frac{\partial^2 {\rm ln} \mathcal{L}(\hat{x}|L, N_{\rm HI})}{\partial \hat{x}^2}\Bigr|_{\hat{x}=\avg{x_{\rm HI}}},
\eeq
where the integration over $\hat{x}$, weighted by $\mathcal{L}(\hat{x})$, takes care of the ensemble averaging. We can rearrange the order of integration and rewrite this as:
\begin{align}\label{eq:lnderiv_two)b}
-\left\langle \frac{\partial^2 {\rm ln} \mathcal{L}(\hat{x})}{\partial \hat{x}^2} \right\rangle \Bigr|_{\hat{x}=\avg{x_{\rm HI}}} &= -\int dL dN_{\rm HI} \frac{dP_L}{dL} \frac{dP_{\rm NHI}}{dN_{\rm HI}} \int d\hat{x}
\mathcal{L}(\hat{x}|L,N_{\rm HI}) \frac{\partial^2 {\rm ln} \mathcal{L}(\hat{x}|L, N_{\rm HI})}{\partial \hat{x}^2}\Bigr|_{\hat{x}=\avg{x_{\rm HI}}} \nonumber \\
& = \int dL dN_{\rm HI} \frac{dP_L}{dL} \frac{dP_{\rm NHI}}{dN_{\rm HI}} \frac{1}{{\rm var}\left[\avg{x_{\rm HI}} | L, N_{\rm HI}\right]}.
\end{align}

Although the notation is a little cumbersome, we arrive at a simple result. The {\em non-conditional} Fisher matrix, $F_{\avg{x_{\rm HI}},\avg{x_{\rm HI}}}$ -- i.e., the inverse variance of the neutral fraction estimate -- is a weighted-average of the conditional neutral fraction inverse variance, where the weighting is over the bubble size and DLA column density PDFs. That is, our final expression is:
\beq\label{eq:fisher_xx_final}
F_{\avg{x_{\rm HI}}, \avg{x_{\rm HI}}} = \int dL dN_{\rm HI} \frac{dP_L}{dL} \frac{dP_{\rm NHI}}{dN_{\rm HI}} \frac{1}{{\rm var}\left[\avg{x_{\rm HI}} | L, N_{\rm HI}\right]}.
\eeq

In summary, our procedure is to first compute the conditional Fisher matrices of Eq.~\ref{eq:fisher_conditional} and Eq.~\ref{eq:fisher_cond_def} for many different values of $L, N_{\rm HI}$, spanning the full range of their PDFs. We then invert each conditional Fisher matrix, using Eq.~\ref{eq:varx_cond} to determine the inverse variance of the neutral fraction estimate for that $L, N_{\rm HI}$. Finally, we average together the results using Eq.~\ref{eq:fisher_xx_final} to weight by the PDFs of $L, N_{\rm HI}$. The final error bar follows simply from this equation as:
\beq\label{eq:error_x_final}
\sigma_{\avg{x_{\rm HI}}} = \left[N_{\rm GRB}(z_{\rm s}) F_{\avg{x_{\rm HI}},\avg{x_{\rm HI}}}\right]^{-1/2}, 
\eeq
where the $N_{\rm GRB}(z_{\rm s})$ factor accounts for the number of independent GRB sightlines in the redshift bin.

\section*{Appendix B}\label{sec:append_b}

Here we show how sensitive our results are to variations around the some of the assumptions in our modeling. First we consider how the results depend on our treatment of the lower limit of integration in our damping wing computations (Eq.~\ref{eq:taudw}). Recall that throughout the text we took each GRB to reside at the center of its redshift bin (at redshift $z_{\rm s}$), and computed the damping wing using the average neutral fraction at bin center, truncating the integral at the edge of each redshift bin, $z_{\rm s} - 0.5$. This neglects redshift evolution across the bin and, more importantly, the presence of neutral gas at redshifts below the lower redshift end of the bin in question. While this estimate is imperfect, it simplifies the calculations and is a reasonable approximation given that the damping wing is dominated by relatively nearby gas in the IGM.

\begin{figure}[htpb]
\bc
\includegraphics[width=0.45\textwidth]{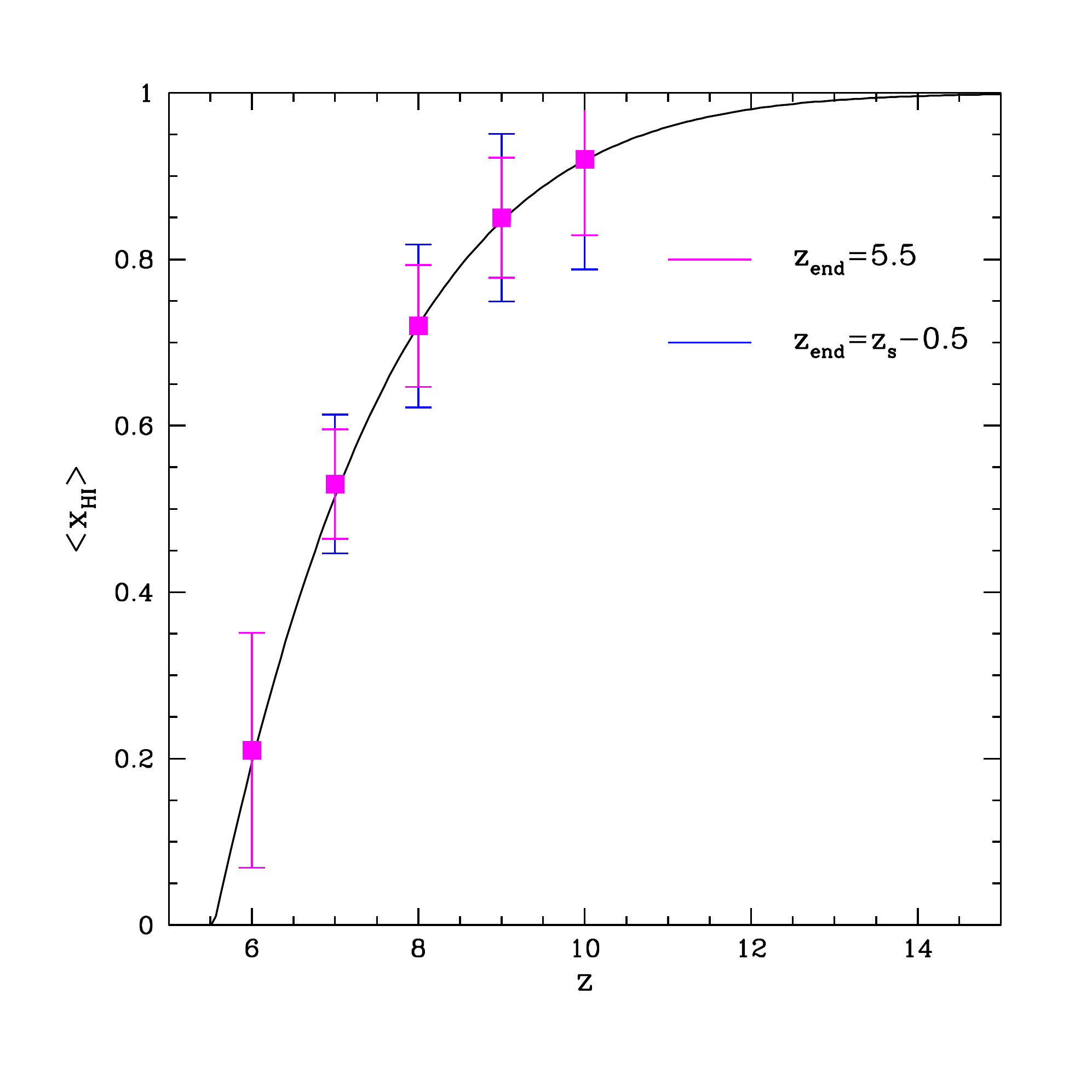}
\caption{Error bar forecasts for different choices of $z_{\rm end}$. The blue points and error bars show our fiducial model assumption for the lower redshift limit on the damping wing integral in Eq.~\ref{eq:taudw}. In this case, $z_{\rm end} = z_{\rm s} - 0.5$: that is, we adopt the average neutral fraction at bin center and integrate down to the lower edge of the bin. The magenta points instead adopt the average neutral fraction at bin center but integrate all the way down to $z_{\rm end}=5.5$. We adopt the fiducial GRB redshift distribution with 20 $z \geq 6$ GRBs. Note that the two cases coincide for the $z_{\rm s}=6$ bin. The results are not strongly sensitive to the treatment of $z_{\rm end}$.}
\label{fig:zend_dep}
\ec
\end{figure}

As one simple test of the impact of this approximation, we simply adjust the lower limit of the redshift integration from $z_{\rm end}-0.5$ to $z_{\rm end}=5.5$, while keeping $\avg{x_{\rm HI}}$ fixed at its bin center value. This latter redshift value coincides with the edge of the lowest redshift bin in our analysis and is close to the end of reionization in the model. This case deliberately exaggerates the impact of neutral gas in the IGM below the edge of each redshift bin, especially in the highest redshift bins (while it gives identical results for the lowest redshift bin). We then follow the procedure in \S \ref{sec:forecasts} to compute the error bars on the ionization history for this alternate choice of $z_{\rm end}$. The results of this test are shown in Figure~\ref{fig:zend_dep}; the error bars tighten slightly in this case especially in the highest redshift bin. However, the difference is relatively small and the case with $z_{\rm end}=5.5$ deliberately exaggerates the effect of the neutral gas below the bin edge. Hence our results are not strongly impacted by the choice of $z_{\rm end}$.

\begin{figure}[htpb]
\bc
\includegraphics[width=0.45\textwidth]{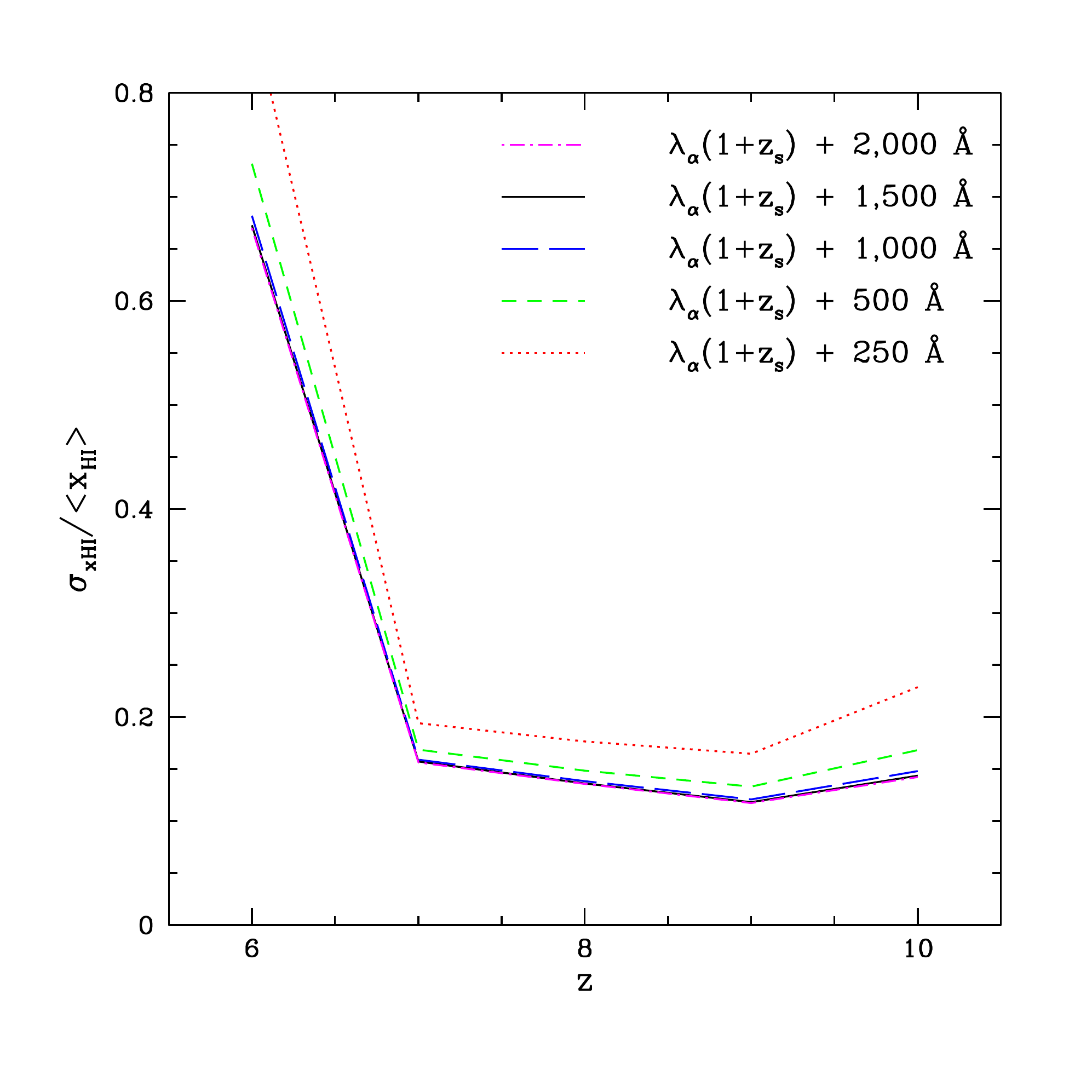}
\caption{Fractional error bars versus redshift for different choices of the maximum wavelength entering into the Fisher matrix computations of Eq.~\ref{eq:fisher_conditional}. Again, the figure shows variations for our fiducial GRB redshift distribution. The text adopts $\lambda_{\rm max} = \lambda_\alpha (1 + z_{\rm s}) + 1,500 \Ang$, but the results are insensitive to this choice.}
\label{fig:almax_dep}
\ec
\end{figure}

A second choice relates to the range of wavelengths included in our Fisher matrix calculations (Eq.~\ref{eq:fisher_conditional}). In the computations in the text we include the wavelength range between Ly-$\alpha$ at the source redshift, at observed wavelength $\lambda_{\rm obs} = \lambda_\alpha (1+z_{\rm s})$, and out to $1,500 \Ang$ redward of this wavelength (independent of $z_s$). In Figure~\ref{fig:almax_dep} we consider variations around this choice. Specifically, for the case of our fiducial GRB redshift distribution we compute the marginalized fractional error on the neutral fraction in each redshift bin, including a range of different choices for the long-wavelength cut-off. In each bin the results converge provided $\gtrsim 500 \Ang$ redward of Ly-$\alpha$ at the source redshift are included (as might also be expected based on Figures~\ref{fig:examp_dwing} and \ref{fig:fisher_derivs}), justifying our fiducial choice of $1,500 \Ang$. 

\begin{figure}[htpb]
\bc
\includegraphics[width=0.45\textwidth]{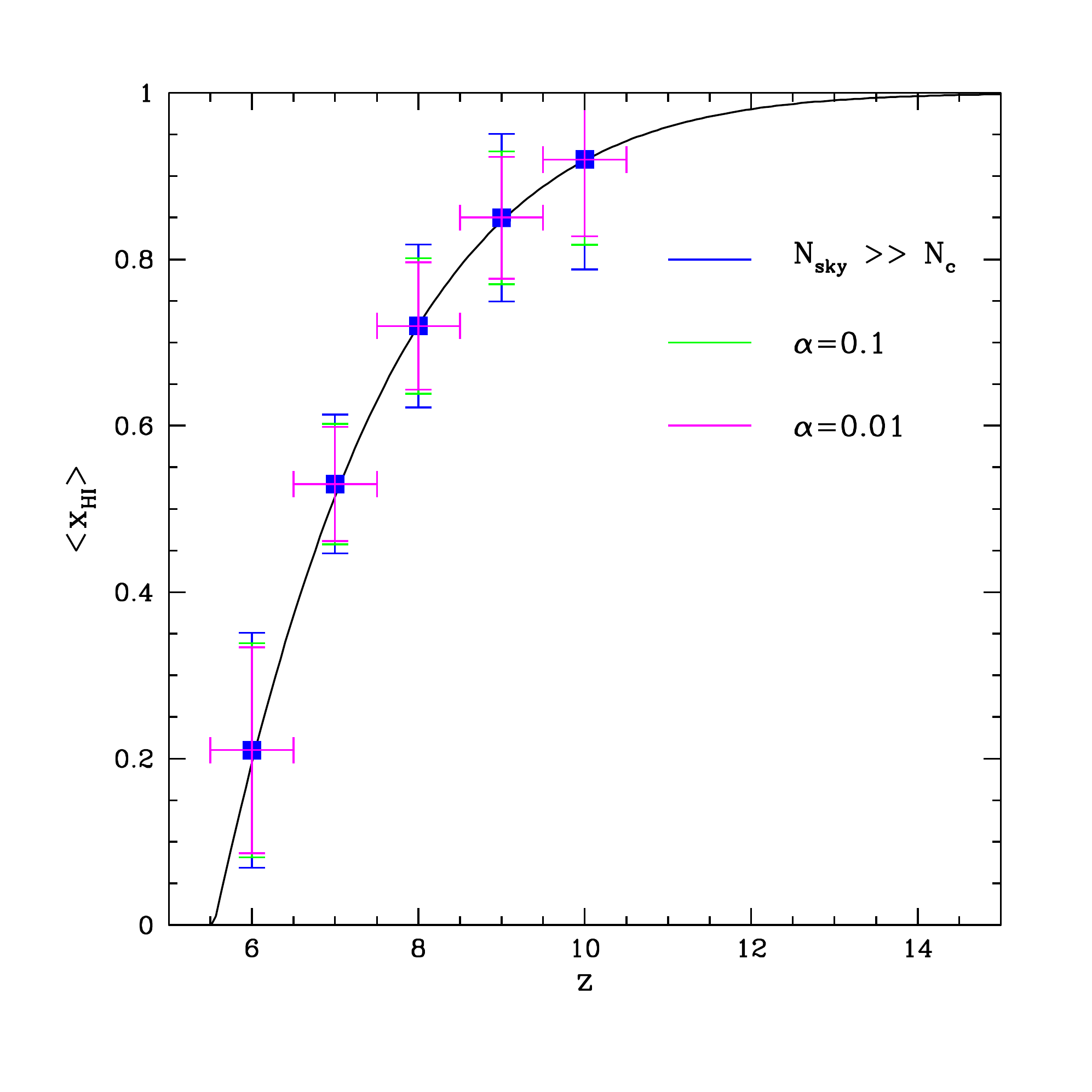}
\caption{Error bar forecasts for different sky background to source count ratios. In each case, the SNR at the continuum is fixed at SNR=20, but the blue points and error bars take the sky background dominated limit adopted in the main text, while the green error bars assume the sky background is $0.1$ times the source flux density, and the magenta errors give the case where the sky background is 100 times smaller than the source. In each case, we assume the fiducial GRB redshift distribution. The results are not sensitive to the assumptions here, although the errors are slightly smaller in the source-dominated case.}
\label{fig:alpha_dep}
\ec
\end{figure}

Finally, the calculations in the text assume sky background limited measurements, which is appropriate for the case of ground-based follow up spectroscopic observations. As discussed in \S \ref{sec:snr_dep}, space-based observations with the JWST will instead be in the source-dominated regime. At fixed SNR, the transmission variance in the source-dominated case will be slightly smaller in absorbed regions than in the sky-dominated regime (see Eq.~\ref{eq:var_flux}). This slightly improves the ability to measure the transmission profile close to Ly-$\alpha$ at the source redshift, which helps in determining the bubble size and column density parameters. This then leads to less strong parameter degeneracies than in the sky-dominated case of Figures~\ref{fig:param_degen_z6}-\ref{fig:param_degen_z10}, and slight improvements on the marginalized neutral fraction error forecasts. Figure~\ref{fig:alpha_dep} compares the sky dominated case with alternate scenarios where the sky flux is $0.1$ and $0.01$ times the source flux. The improvements are relatively small, and so we expect rather similar constraints at fixed SNR in each of the sky and source dominated cases. 

\end{document}